 \documentclass[final,3p,times,twocolumn]{elsarticle}

\usepackage{amssymb}

\usepackage{lineno}
\usepackage{geometry}   
\usepackage{multirow}
\usepackage{subfig}
\usepackage{graphicx}
\usepackage{amsmath}
 \usepackage{amsthm}
\usepackage{mathrsfs}
\usepackage{color}
\journal{Astroparticle Physics}

\begin{document}

\begin{frontmatter}
\title{First Constraints on the Ultra-High Energy Neutrino Flux from a Prototype Station of the Askaryan Radio Array}



\author[6]{P. Allison}
\author[8]{J. Auffenberg}
\author[7]{R. Bard}
\author[6]{J. J. Beatty}
\author[4,5]{D. Z. Besson}
\author[3]{C. Bora}
\author[9]{C.-C. Chen}
\author[9]{P. Chen}
\author[6]{A. Connolly}
\ead{connolly@physics.osu.edu}
\author[1]{J. P. Davies}
\author[8]{M. A. DuVernois}
\author[2]{B. Fox}
\author[2]{P. W. Gorham}
\author[11]{K. Hanson}
\author[2]{B. Hill}
\author[7]{K. D. Hoffman}
\author[6]{E. Hong}
\author[9]{L.-C. Hu}
\author[12]{A. Ishihara}
\author[8]{A. Karle}
\author[8]{J. Kelley}
\author[3]{I. Kravchenko}
\author[10]{H. Landsman}
\author[8]{A. Laundrie}
\author[9]{C.-J. Li}
\author[9]{T. Liu}
\author[8]{M.-Y. Lu}

\author[7]{R. Maunu}
\author[12]{K. Mase}
\author[11]{T. Meures}
\author[2]{C. Miki}
\author[9]{J. Nam}

\author[1]{R. J. Nichol}
\author[10]{G. Nir}
\author[11]{ A. O'Murchadha}
\author[6]{C. G. Pfendner}

\author[14]{K. Ratzlaff}
\author[7]{M. Richman}
\author[2]{B. Rotter}
\author[8]{P. Sandstrom}
\author[13]{D. Seckel}
\author[3]{A. Shultz}
\author[4]{J. Stockham}
\author[4]{M. Stockham}
\author[5]{M. Sullivan}
\author[7]{J. Touart}
\author[9]{H.-Y. Tu} 
\author[2]{G. S. Varner}
\author[12]{S. Yoshida}
\author[14]{R. Young}

\address[6]{Dept. of Physics and CCAPP, The Ohio State University, 191 W. Woodruff Ave., Columbus, OH 43210, USA}
\address[7]{Dept. of Physics, University of Maryland, College Park, MD 20742, USA}
\address[4]{Dept. of Physics and Astronomy, University of Kansas, 1251 Wescoe Hall Dr., Lawrence, KS 66045, USA}
\address[5]{Moscow Engineering and Physics Institute, 31 Kashirskaya Shosse, Moscow 115409, Russia}
\address[3]{Dept. of Physics and Astronomy, University of Nebraska-Lincoln, 855 N 16th Street, Lincoln, NE 68588, USA}
\address[9]{Dept. of Physics, Grad. Inst. of Astrophys.,\& Leung Center for Cosmology and Particle Astrophysics, National Taiwan University, No. 1, Sec. 4, Roosevelt Road, Taipei 10617, Taiwan (R.O.C.)}
\address[1]{Dept. of Physics and Astronomy, University College London, Gower Street, London WC1E 6BT, United Kingdom}
\address[2]{Dept. of Physics and Astronomy, University of Hawaii-Manoa, 2505 Correa Rd., Honolulu, HI  96822, USA}
\address[8]{Dept. of Physics and Wisconsin IceCube Particle Astrophysics Center, University of Wisconsin-Madison, 222 W. Washington Ave, Madison, WI 53706, USA}
\address[10]{Department of Particle Physics and Astrophysics, Weizmann Institute of Science, Rehovot, 76100, Israel}
\address[11]{Service de physique des particules \'{e}l\'{e}mentaires, Universit\'{e} Libre de Bruxelles, 	CP230, boulevard du Triomphe, 1050 Bruxelles, Belgium}
\address[12]{Dept. of Physics, Chiba University, 1-33, Yayoi-cho, Inage-ku, Chiba-shi, Chiba 263-8522, Japan}
\address[13]{Dept. of Physics and Astronomy, University of Delaware, 104 The Green, Newark, DE 19716, USA}
\address[14]{Instrumentation Design Laboratory, University of Kansas, 1251 Wescoe Drive, Lawrence, KS 66045, USA}

\begin{abstract}
The Askaryan Radio Array (ARA) is an ultra-high energy ($>10^{17}$~eV) cosmic neutrino detector in phased construction near the south pole. 
ARA  searches for radio Cherenkov emission from particle cascades induced by neutrino interactions in the ice using radio frequency antennas ($\sim150-800$~MHz) deployed at a design depth of 200~m in the Antarctic ice.
A prototype ARA Testbed station was deployed at $\sim30$~m depth in the 2010-2011 season and the first three full ARA stations were deployed in the 2011-2012 and 2012-2013 seasons. 
We present the first neutrino search with ARA using data taken in 2011 and 2012 with the ARA Testbed and the resulting constraints on the neutrino flux from $10^{17}-10^{21}$~eV.
\end{abstract}

\begin{keyword}
GZK effect
\sep UHE neutrinos
\sep radio Cherenkov


\end{keyword}

\end{frontmatter}

\section{Introduction}

The Askaryan Radio Array (ARA) aims to measure the flux of ultra-high energy (UHE) neutrinos above $10^{17}$~eV.
 While UHE neutrinos are so far undetected, they are expected both directly from astrophysical sources and as decay products from 
 the GZK process \cite{Greisen:1966jv,Zatsepin:1966jv}, as first pointed out by  
 Berezinsky and Zatsepin~\cite{Berezinsky:1969zz,Berezinsky:1970}.  The GZK process describes the interactions between
 cosmic rays and cosmic microwave and infrared background photons above a $\sim10^{19.5}$~eV threshold.

The interaction of a UHE neutrino in dense media induces an electromagnetic shower which in turn creates impulsive radiofrequency (RF) Cherenkov emission via the Askaryan effect~\cite{Askaryan:1962,Askaryan:1965,Zas:1991jv,Gorham:2000ed,Saltzberg:2000fk,Gorham:2004ny,Gorham:2006fy}.
In radio transparent media, these RF signals can then be observed by antenna arrays read out with $\sim$~GHz sampling rates.

Currently, the most stringent limits on the neutrino flux above $\sim10^{19}$~eV have been placed by the balloon-borne ANITA experiment sensitive to impulsive radio signals from the Antarctic ice sheet~\cite{Gorham:2010kv,Gorham:2010xy}.
Below $10^{19}$~eV, the best constraints on the neutrino flux currently come from the IceCube experiment, a $1~\rm{km}^3$ array of photomultiplier tubes in the ice at the south pole using the optical Cherenkov technique~\cite{Aartsen:2013dsm}.  
IceCube has recently reported the first extraterrestrial high energy neutrino flux, which extends up to $\sim10^{15}$~eV.  This is two orders of magnitude lower energy than ARA's energy threshold~\cite{Aartsen:2013jdh,Aartsen:2014gkd}.

Due to the $\sim 1$~km radio attenuation lengths in ice ~\cite{Allison:2011wk, Barwick:2005zz}, radio arrays have the potential to view the $100$s of $\rm{km}^3$ of ice necessary to reach the sensitivity to detect $\sim10$ events per year from expected UHE neutrino fluxes.
The first radio array in ice to search for UHE neutrinos, RICE, was deployed along the strings of the AMANDA detector,
an IceCube predecessor, and placed competitive limits on the UHE neutrino flux between $10^{17}$ and $10^{20}$~eV~\cite{Kravchenko:2011im}.
Next-generation detectors are under construction aiming to reach the $100$s of $\rm{km}^3$ target volume of ice.
The Askaryan Radio Array (ARA)~\cite{Allison:2011wk} is one such detector being deployed in the ice at the south pole and the 
first physics results from a prototype station of this detector are presented in this paper.
Another experiment with similar aims, ARIANNA, is currently being developed on the surface of the Ross Ice Shelf in Antarctica~\cite{Gerhardt:2010js}.

\begin{figure}[t]
  \centering
  \includegraphics[width=0.5\textwidth]{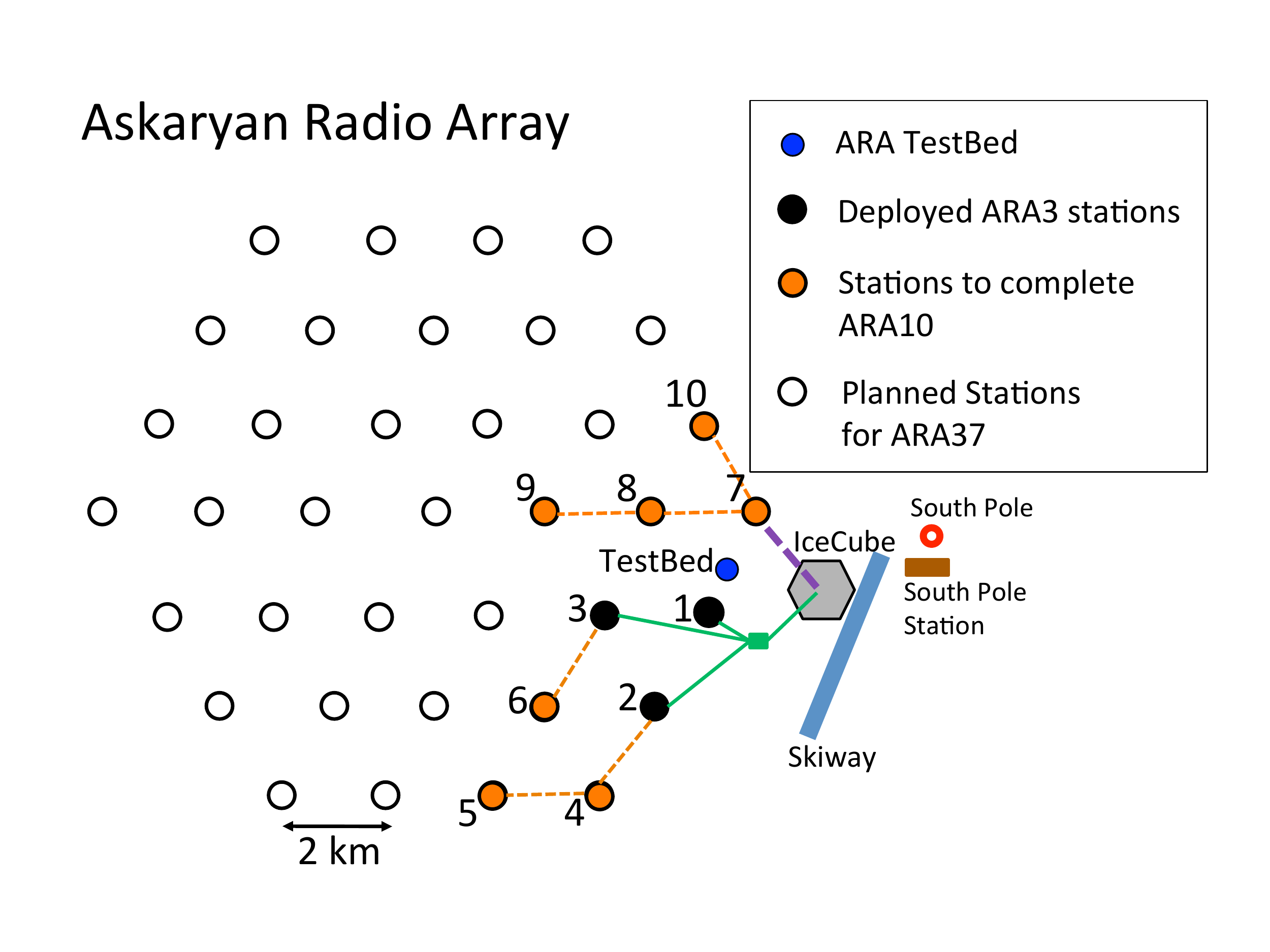}
  \caption{Diagram showing the layout of the proposed ARA37 array, with the location of the Testbed and the first three deployed
  deep stations highlighted in blue and black respectively, and proposed stations for the next stage of deployment, ARA10, highlighted in orange.}
  \label{fig:ara37}
 \end{figure}

ARA aims to deploy 37 stations of antennas at 200~m depth spanning $100~\rm{km}^2$ of ice as shown in Fig.~\ref{fig:ara37}.
A design station consists of eight horizontally polarized (HPol) and eight vertically polarized (VPol) antennas at depth and four surface antennas for background rejection and cosmic ray detection via the geomagnetic emission in the atmosphere.
The 200~m design depth was chosen because it is below the firn layer, 
where the index of refraction varies with depth due to the gradual compacting of snow into ice down to $\sim150$~m depth.
The trigger and data acquisition are handled by electronics at the surface of the ice at each station.

To date, one ARA prototype Testbed station and three full stations have been deployed in the ice. 
The Testbed station was deployed at a depth of $\sim30$~m in the 2010-2011 drilling season.
The first full station, A1, was deployed at a depth of 100~m in the 2011-2012 drilling season.
The next two stations, A2 and A3, were deployed at the 200~m design depth during the 2012-2013 season.
At the time of publication, station A2 and A3 are operational while A1 is under repair.

This paper presents three complementary analyses using
data taken with the Testbed station. 
The first two analyses use a series of cuts to reject background signals in favor of neutrino events.
The third analysis is template-based and searches for unique impulsive signals after correlating events.

  \begin{figure}[t]
  \centering
  \includegraphics[width=0.45\textwidth]{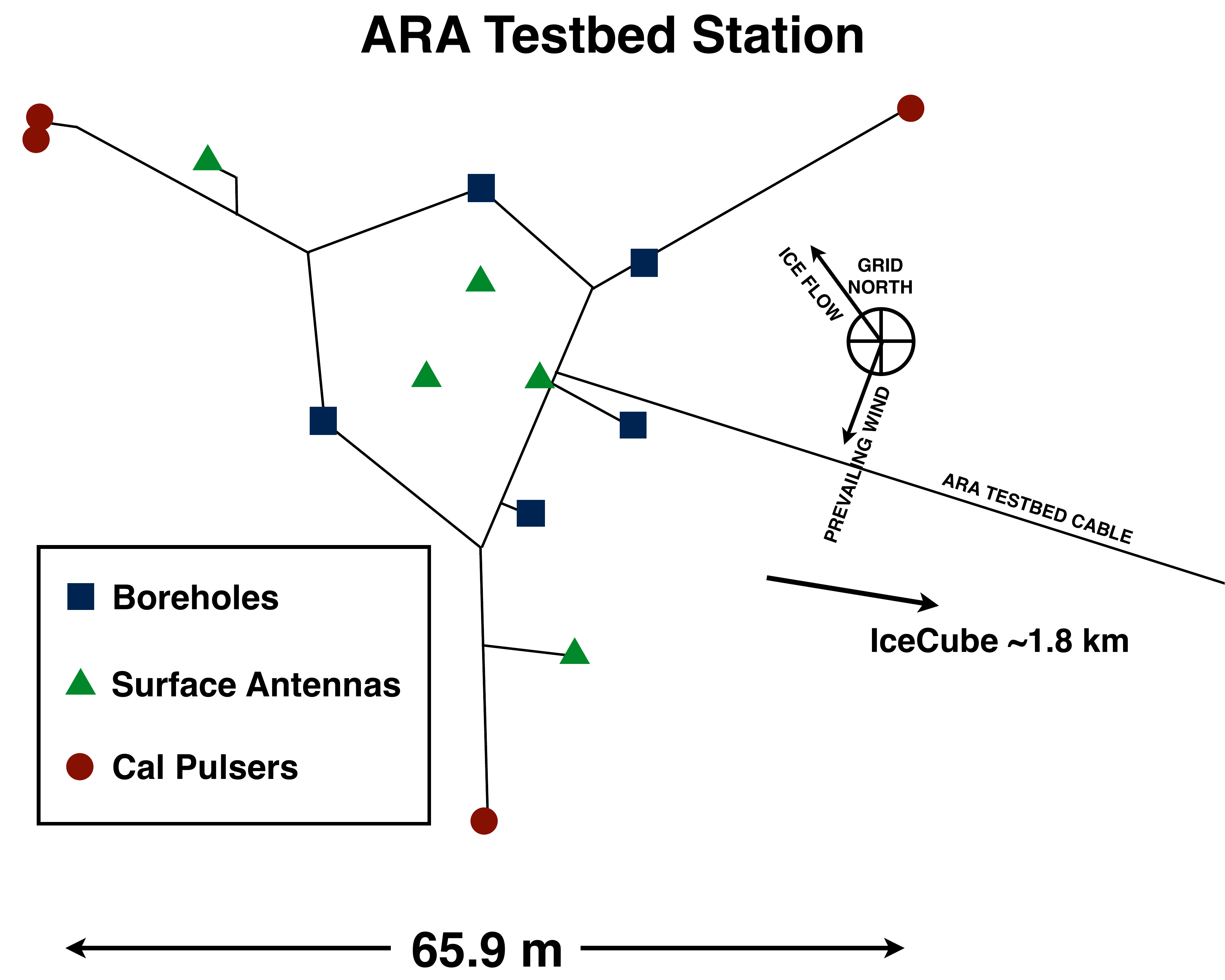}
  \caption{Schematic of the ARA Testbed station.}
  \label{fig:testbedstation}
 \end{figure}

 \section{Testbed}
 The ARA prototype Testbed station differs from the layout of the design stations for the full array.  A more complete description of the design and operation of the Testbed station can be found in~\cite{Allison:2011wk}.

 \begin{table}[t]
\begin{tabular}{ l | c | c | c | c} \hline
\multirow{2}{*}{Hole} & x & y   & Type, Pol & Depth   \\ 
 & (m) & (m) & & (m) \\ \hline
\multirow{2}{*}{BH 1} &  \multirow{2}{*}{-8.42} & \multirow{2}{*}{-4.40} & BSC, H & 20.50  \\ 
 & & & Bicone, V & 25.50 \\ \hline
\multirow{2}{*}{BH 2} &  \multirow{2}{*}{-0.42} & \multirow{2}{*}{-11.13} & BSC, H & 27.51  \\ 
& & & Bicone, V & 22.51 \\ \hline
\multirow{2}{*}{BH 3} & \multirow{2}{*}{9.22} & \multirow{2}{*}{-6.15} &  BSC, H & 22.73   \\ 
& & & Bicone, V & 27.73 \\ \hline
\multirow{2}{*}{BH 5} &  \multirow{2}{*}{3.02} & \multirow{2}{*}{10.41} & BSC, H & 30.56   \\ 
& & & Bicone, V  & 25.56 \\ \hline
\multirow{2}{*}{BH 6} & \multirow{2}{*}{-9.07} & \multirow{2}{*}{3.86} & QSC, H & 26.41  \\
& & & QSC, H & 30.41 \\  \hline
\multirow{2}{*}{S1} & \multirow{2}{*}{-2.48} & \multirow{2}{*}{-1.75} & Discone, V   & 1.21  \\ 
& & & Batwing, H & 2.21 \\ \hline
S2 & 4.39 & -2.41 & Batwing (H) & 1.19 \\ \hline
S3 & 1.58 & 3.80 & Discone (V) & 1.19    \\  \hline
S4 & & & Fat Dipole (H) & \\ \hline
\multirow{2}{*}{Cal 1} & \multirow{2}{*}{-23.18} & \multirow{2}{*}{17.90} & H  & 17.50   \\ 
& & & V & 22.50 \\ \hline
\multirow{2}{*}{Cal 2}  & \multirow{2}{*}{-2.25} & \multirow{2}{*}{-29.81} & H  & 34.23    \\ 
& & & V & 29.23 \\ \hline
\multirow{2}{*}{Cal 3}  &  {27.67} &  {13.57} & H  & 1.13  \\ 
& 28.69 & 12.35 & V & 1.13 \\ \hline
\end{tabular}
\caption{\label{tab:positions} Types and positions of antennas as deployed in the ARA Testbed.   See the text for the description
of antenna types.}
\end{table}

Table~\ref{tab:positions} summarizes the antenna types and deployed positions in the Testbed which are depicted in
 Fig.~\ref{fig:testbedstation}. 
Here we use 
the Testbed-centric coordinate system with
the origin at the southeast corner of the DAQ box on the surface of ice, $+\hat{x}$ pointing along the direction of ice flow
and the $\hat{x}-\hat{y}$ plane tangent to the earth's geoid shape at the surface.

As with the deep stations, the Testbed antennas deployed in boreholes were designed to be broadband, with a mixture of HPol and VPol,  subject to the constraint that they must fit down the $\sim15$~cm diameter hole in the ice.  
For VPol, a wire-frame hollow-center biconical design was chosen with an annular-shaped feed with the string cable running through the center.  
These ``Bicones'' have a bandwidth of 150-850~MHz, and four were deployed in boreholes and two near the surface.    
For HPol, two designs were used in the Testbed, the bowtie-slotted cylinder (BSC) and the quad-slotted-cylinder (QSC).  The BSCs were used in four borehole antennas and a pair of QSC's in the fifth borehole.  Among the 10 borehole antennas, only the eight that are not QSC antennas are used for the trigger.
QSC antennas were deployed in the Testbed to test the antenna design before deploying them in the deep stations.
Of the three analyses presented in this paper, only the Template-Based analysis utilized the two QSC antennas. 

Larger antennas were deployed at the surface.  Two discone antennas (VPol) and two Batwings (HPol) were deployed 1-2~m
from the surface.  Additionally, two fat dipoles with a bandwidth of 30-300~MHz were deployed within a meter of the surface to 
assess the feasibility of detecting geosynchotron RF emission from cosmic rays, which has a lower frequency content than the
Askaryan emission expected from neutrinos.

Within 1~m of the antennas, a filter and $\sim$40~dB low noise amplifier (LNA) prepare the signal for transmission to the electronics box at the surface.
A notch filter at 450~MHz removes the south pole communications from the Land Mobile Radio handheld UHF systems.
A bandpass filter sits just after each antenna and blocks power outside of our 150~MHz to 850~MHz band before amplification.
The filtered signal in each antenna is then input to a low noise amplifier and transmitted to the surface.
At the surface, a second stage $\sim$40~dB amplifier boosts the signals before they are triggered and digitized.   
After arriving at the electronics box at the surface, the signals are split into a path to the trigger, which determines when a signal is to be stored, and another path to the digitizer, which reads out the waveforms.

There are two different trigger modes in the Testbed, an RF trigger and a software trigger.
An event passes the RF trigger when the output of a tunnel diode, a few-ns power integrator of the waveforms from each antenna
reaching the trigger path,
exceeds 5-6 times the mean noise power in three out of the 8 borehole antennas within a 110~ns coincidence window.
Due to the differences in responses between channels, each antenna has a slightly different power threshold but 
they may be adjusted together to obtain different trigger rates.
The software trigger causes an event to be recorded every second to monitor the RF environment.

Once the station has triggered, the digitization electronics, which are descended from those developed for ANITA ~\cite{Varner:2007zz}, process the waveforms and output them to storage.
Here, in the digitizer path, the signal undergoes an analog-to-digital conversion using the LAB3 RF digitizer~\cite{Varner:2007zz}, and is stored in a buffer (in the Testbed, the buffer was trivially one event deep). 
The signals from the ``shallow" antennas are sampled at 1 GHz, while the signals from the eight borehole antennas were sampled twice, with a time offset of 500 ps for an effective sampling rate of 2 GHz.
The digitized waveforms are $\sim$250~ns long and are centered within approximately 10 ns of the time the station triggered.

Three calibration pulser VPol and HPol antenna pairs were installed at a distance of $\sim$30~m from the center of the Testbed array to provide {\it in situ} timing calibration and other valuable cross checks related to simulations and analysis.
An electronic pulser in the electronics box produces a $\sim$250~ps broadband impulsive signal at a rate of 1~Hz.
This pulser is connected to one of the three calibration pulser antenna pairs and can transmit from either the VPol or HPol antenna in each pulser borehole.
Having multiple calibration pulser locations provides a cross check for the timing calibrations of each channel. 
Also, the observation, or non-observation, of the constant pulse rate by the station provides an estimate of its livetime.

For the Testbed, an event filter selects one event from every ten events at random to be transmitted to the North by satellite and the remaining data is stored locally and hand-carried during the following summer season.
For the other ARA stations, this filter is now optimized to select events that exhibit a causal trigger sequence and thus are more likely to be events of interest.

\section { Simulations }
\label{sec:simulations}
There are many simulations based at different institutions used to model ARA.
The official simulation program for ARA is called AraSim, and is the one used for the Interferometric Map Analysis and the Coherently Summed Waveform Analysis presented in this paper.  
AraSim draws on ANITA heritage~\cite{Gorham:2008dv}, but much of the program was custom developed for ARA.
An alternate simulation program (RA-RA) is used for the Template-Based Analysis.  
One distinguishing feature of RA-RA is that it does not model thermal noise from first principles, but instead inserts simulated neutrino signals on top of measured software-triggered waveforms.  
AraSim does not take this approach because its trigger model requires generating noise waveforms much longer than the length of a waveform recorded in an event.  

\subsection{AraSim}
AraSim generates neutrino events independent of each other, with uniformly distributed neutrino directions and with interaction point locations chosen with a uniform density in the ice. 
For computational ease, neutrinos are generated within a 3-5~km radius around the center of a single station for neutrino energies 
from $E_{\rm{\nu}}=10^{17}$-$10^{21}$~eV, with the larger radii used for higher energies.  For simulating multiple stations, 
neutrino interactions are generated up to 3-5~km beyond the outermost stations.
The energy of the simulated neutrino event can be set to a fixed value or selected from a chosen energy spectrum.
Each event is given a weight equal to the probability that the neutrino  would have reached the interaction point without being absorbed in the Earth using the energy-dependent cross sections in~\cite{Connolly:2011vc}.  Inelasticity distributions are also taken from~\cite{Connolly:2011vc}
and used to allocate the energy of the hadronic shower and any electromagnetic shower. 

In addition to the showers from the primary interaction, AraSim considers any secondary interactions from $\mu$ or $\tau$ leptons that are generated from neutrino-ice charged current interactions.
AraSim calculates the total energy of the primary showers, hadronic and electromagnetic if there is one, and the energy of the secondary showers and generates the RF signal from the interaction among them that produces the most shower energy.

The primary shower comes from the initial neutrino-ice interaction.
The energy for each electromagnetic and hadronic shower from the primary interaction is obtained from inelasticity distributions from~\cite{Connolly:2011vc}.
Electromagnetic and hadronic shower energies for the secondary showers from any $\mu$ or $\tau$ leptons are calculated from interaction probability tables obtained from the MMC particle generation code~\cite{MMC:2004}.
AraSim generates the RF emission from the interaction that produced the most shower energy and then progresses to the trigger simulation.

Since the observed RF emission depends upon the angle relative to the Cherenkov angle at which a given antenna views the event,
the path of the signal through the ice is needed.
To find the signal path, we use a depth-dependent index of refraction fit to an exponential model fitted to data from the RICE experiment~\cite{2004JGlac..50..522K}. 
A ray-tracing algorithm then finds the (zero, one or two) path solutions between the chosen interaction point and the position of each antenna taken from the measured coordinates at deployment.  If there are two solutions, we model both received impulses for that event. 

In AraSim a in-ice ray tracing code named RaySolver drives multiple source to target ray-trace solutions.
From RaySolver, we obtain the travel distance and time from the shower to each antenna and also the polarization of the signal and the receiving angle at the antenna so that we can 
apply antenna responses to the signal.

RaySolver has a multi-step processes to optimize the computation time.
For the first step, it uses an equation, which is driven from Snell's law, to find the launch angle $\theta_0$ at the source location.
This equation 
is not analytically solvable but can be solved numerically (see~\ref{sec:appen_raytracing}):

\begin{align}
\label{eq:raytracing}
&\ln \left(  \frac{\sqrt{\left( n^2-\sigma_0^2n_0^2\right) \left( A^2-\sigma_0^2n_0^2 \right)} + An-\sigma_0^2 n_0^2}{n-A}\right) \nonumber \\
& - \ln \left(  \frac{\sqrt{\left( n_0^2-\sigma_0^2n_0^2\right) \left( A^2-\sigma_0^2n_0^2 \right)} + An_0-\sigma_0^2 n_0^2}{n_0-A}\right) \nonumber \\
& + \frac{C\sqrt{A^2-\sigma_0^2 n_0^2}}{\sigma_0 n_0} (x - x_0) = 0 
\end{align}
where $A$ is one of the parameter values from the index of refraction model ($n(z) = A + B e^{C \cdot z}$), $n$ and $n_0$ are index of refraction value at the source and the target respectively, $x-x_0$ is the horizontal distance between the source and the target, and $\sigma_0 = \sin \theta_0$.
The inputs to Equation~\ref{eq:raytracing} are $x-x_0$, $n_0$ and $n$, which are different for each event.
From this equation, RaySolver finds an initial launch angle at the source location that makes the left-hand-side of the equation smaller than $10^{-4}$.
This first semi-analytic approach is much faster than ordinary ``trial and error" method as we don't need to trace the ray step-by-step for multiple trials.
If RaySolver couldn't find the solution with the first semi-analytic method, it uses a ``trial and error" technique to find the solution.
If RaySolver finds a first solution, it moves on to the next possible solution which is either a U-turned (or highly bent) in-ice trace or a surface reflected trace.
It uses only a traditional ``trial and error" method to search for a second solution. 
For each solution, the minimum distance between the ray and the target should be less than the required accuracy parameter which is 0.2~m.

\begin{figure}[t]
  \centering
  \includegraphics[width=0.5\textwidth]{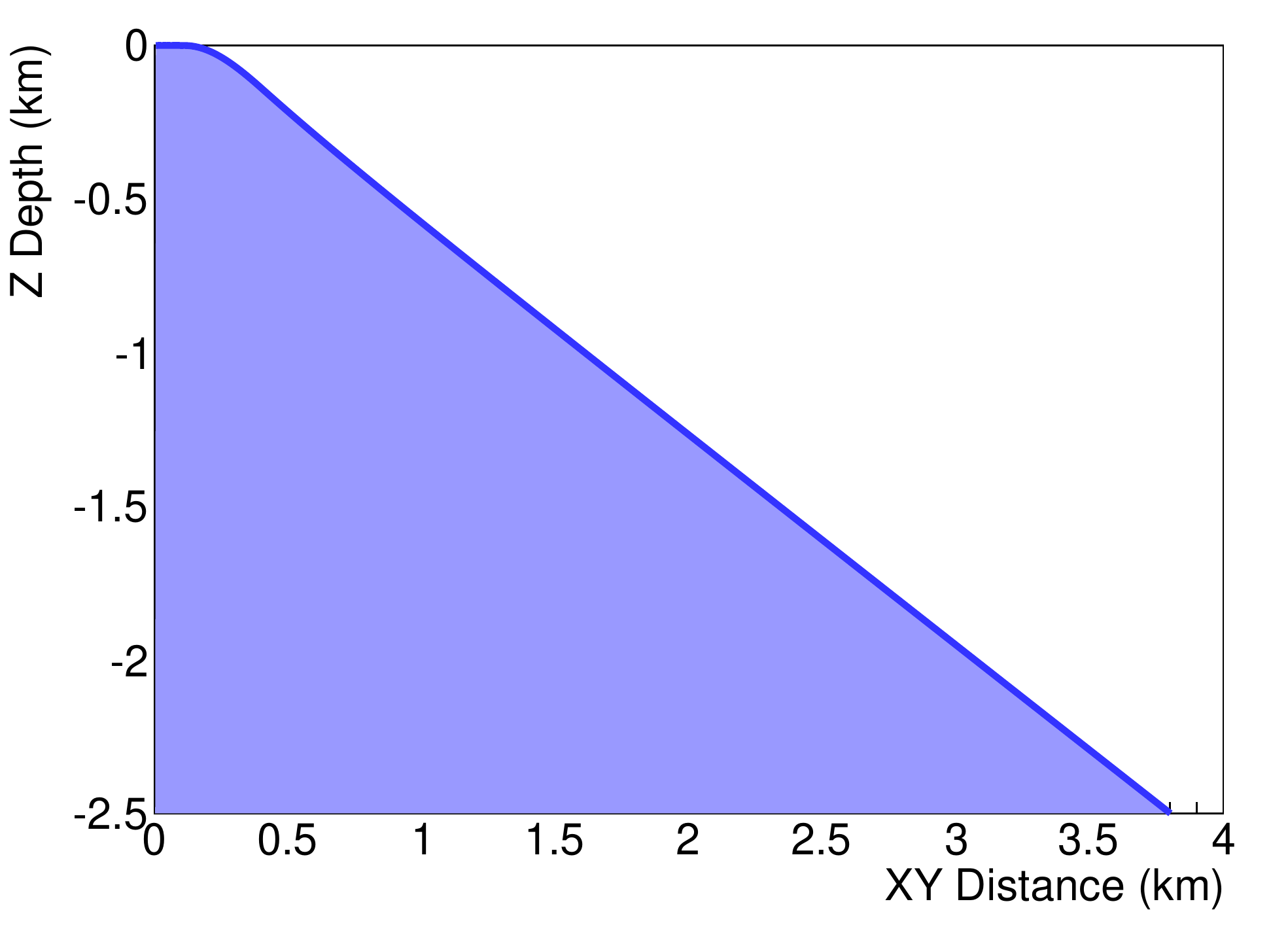}
  \includegraphics[width=0.5\textwidth]{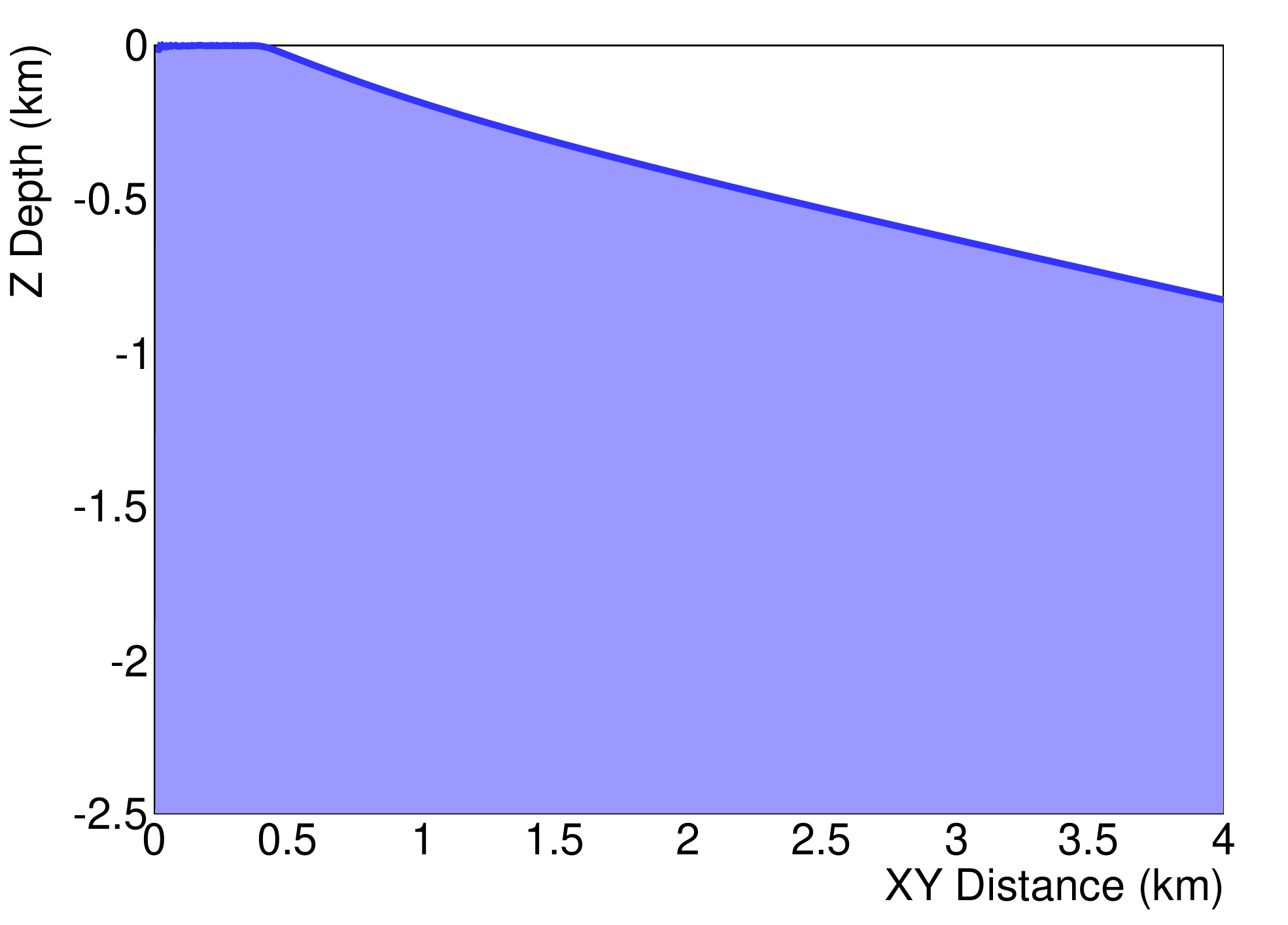}
  \caption{Plot showing the regions with ray-trace solutions for an antenna depth at 25~m (top) and 200~m (bottom). The greater depth allows an antenna at 200m depth to observe a larger volume of the ice.}
  \label{fig:RayTrace}
 \end{figure}

The depth-dependent index of refraction model is a necessary component for the proper simulation of the environment as it causes the signal path to bend in the ice. 
For depths within the firn ($<$150~m), this curvature effect is significant and large regions of the ice beyond $\sim$1~km away have no ray-trace solutions to the antennas as can be seen in Fig.~\ref{fig:RayTrace}.
By increasing the depth of the station from 25~m to the ARA design depth of 200~m (below the firn), the station's effective volume increases by a factor of three at $E_{\rm{\nu}}=10^{18}$~eV and a factor of four at $E_{\rm{\nu}}=10^{19}$~eV.
This increase is due to two factors: (1) the total viewable volume is increased and (2) more neutrinos with down-going or Earth-skimming incident angles produce observable interactions and are less subject to earth absorption. 

Once the event is generated and the ray-trace solutions are found, the RF Cherenkov emission 
is modeled from a custom algorithm that uses a parameterized model of the shower and its RF emission adapted from~\cite{AlvarezMuniz:2010ty}. 
Based on the energy of the neutrino event, AraSim generates a charge excess profile for the shower based on the Greisen function~\cite{Greisen:1956} for the electromagnetic component of the shower and the Gaisser-Hillas function~\cite{Gaisser:1977} for the hadronic component~\cite{jaime}.
Then the vector potential is calculated from a parametrized model~\cite{AlvarezMuniz:2011ya} that requires the charge excess profile and the viewing angle relative to the Cherenkov angle found from the ray-tracing algorithm as inputs.
The time-dependent electric field is then just the time derivative of the vector potential.

This simulation of RF emission does not yet include the Landau-Pomeranchuk-Migdal (LPM) ~\cite{Landau:1953um,Landau:1953gr,Migdal:1956tc} effect, where electromagnetic showers are lengthened due to quantum mechanical interference resulting in a narrowing of the Cherenkov cone.
However, we have accounted for the LPM effect in our limits by applying a scale factor at each energy.
These scale factors are obtained by running AraSim with the fully parametrized RF emission as given in~\cite{AlvarezMuniz:2000fw,AlvarezMuniz:1997sh} and taking the ratio of the effective volume with the LPM effect on to the effective volume with the LPM effect off.
These corrections are at approximately the 20\% level.
Future versions of our RF emission model in AraSim will account for the LPM effect.

After the RF emission is calculated, it is propagated through the ice model to the antennas.
The signals are attenuated in the ice according to~\cite{Barwick:2005zz}, but with no frequency dependence, using  the distance along each ray-trace path through the ice.
The signal arrival times are calculated for each path taking into account the depth-dependent index of refraction. 

After the signals reach the antennas, we model the response of the detector to these received signals.
The antenna angular and frequency responses used were created in NEC2~\cite{NEC2} simulations of similar antenna types.
We modified the environmental condition of NEC2 software in order to simulate the antennas in an ice.
After the received signal is convolved with the antenna response, the response of the RF filters and amplifiers are then applied.
The amplitude response of the filters and amplifiers are taken from lab measurements while the phase response is derived from a Qucs Studio~\cite{QS} model of filters with similar characteristics.
Once the received signal is convolved with the detector response, noise is added to the signal.

The noise modeling in AraSim is derived from 
software triggered events recorded by the Testbed between February and June 2012.  These type of events 
should consist primarily of thermal background.
For each antenna, we compute the frequency spectrum for each software triggered event and find the distribution of amplitudes over all events for each of
512 frequency bins of $\sim$2~MHz width.
In each frequency bin, 
the distribution of these amplitudes is fitted to a Rayleigh function, and the best fit function 
is used by AraSim to generate noise for that antenna for that frequency bin. 
In each bin, an amplitude is randomly selected from the appropriate Rayleigh distribution and a phase is selected from a flat distribution.
Once the amplitudes and phases have been selected for all frequency bins, the spectrum is converted to the time domain to produce a $\sim6~\mu$s noise waveform or one long enough to encompass the arrival times for all ray solutions.  
Since 8192 frequency bins are necessary for generating such a long waveform, interpolation of the fit parameters is performed 
between the 512 frequency bins. 
The time-domain signal waveform is added to this noise waveform at its calculated arrival time at the antenna.
This process is repeated for each antenna and thus the simulated noise level  is at the same level  as the
 the recorded data for each antenna. 

Once the noise has been added, the signal is split into the trigger and digitization paths.
For the trigger path, the time-domain signal is convolved with a model of the tunnel diode power integrator.
This convolved time-domain response is then scanned for excursions above the power threshold.
For the Testbed simulation, the power thresholds were calibrated against RF triggered events for each antenna.
When the trigger finds 3 such excursions among the 8 borehole antennas within a 110~ns window, the event is considered to have triggered.
Once the trigger condition is met, waveforms are read out in 256~ns waveforms just like the data, and written into the same 
format as the data so that the simulated events can be analyzed with identical software.

\subsection{RA-RA}
A second simulation (RA-RA) traces its heritage to the RICE experimental simulation and is entirely independent of the AraSim code. The signal generation is based on the Alvarez-Muniz {\it et al.} parametrization of the Askaryan effect~\cite{AlvarezMuniz:2010ty}, and primarily differs from the standard AraSim package in the way signal is overlaid on background. In this case, forced trigger events sampled over the period corresponding to the data analysis are retained and are considered as representative of the environment against which a signal neutrino event must be discriminated. Correspondingly, neutrinos are simulated without noise, then their signal voltages directly added to the voltage vs. time data from the forced triggers. In addition, the RA-RA simulation models birefringence, which is expected to be a subtle
effect but one that is not modeled in AraSim. 

Even after the signals are overlaid over the waveforms from each force triggered event, there is still an uncertainty in the actual time when the event trigger latches in the waveform. The Testbed electronics have an intrinsic total jitter of 25~ns in this actual time. In the RA-RA simulation, the third antenna of a given polarization with a voltage excursion exceeding $4.5~\sigma$ is assumed to be that channel that satisfied the online 3/8 requirement; the time at which that excursion occurred is assigned a value between -12.5 and 12.5~ns, with a uniform probability for all values in between. 

The advantage of this scheme where signals are added to force triggered waveforms is that it includes all of the backgrounds that are present in the data; the disadvantage is that the trigger simulation is somewhat cruder than in the default AraSim package. However, in so far as the subsequent analysis makes a more stringent requirement of four antennas with at least 
$6~\sigma_{\rm{V}}$ 
(vs. three $4.5~
\sigma_{\rm{V}}$ 
excursions for the trigger simulation), the neutrino detection efficiency is set by the analysis, rather than the trigger requirements, in any case.

\section { Testbed Data Analyses }

This paper presents three complementary  searches for neutrino events in the Testbed.  
The Interferometric Map Analysis aims to reduce backgrounds mainly by assessing the quality of reconstructions produced interferometrically.  
This analysis performs depth-dependent ray tracing through the firn layer and we use the results of this analysis to derive constraints on the neutrino flux at the conclusion of the paper. 
The Coherently Summed Waveform Analysis uses a different reconstruct technique, performing a best fit to time delays derived from coherently summed waveforms.  
This analysis is complementary to the first one in that it has $\sim30\%$ higher analysis efficiency but a $\sim10\%$ lower livetime, leading to very similar limits as those set by the first analysis. 
The third analysis, the Template-Based Analysis, performs correlations between events and searches for any producing a unique pattern of waveforms.  In contrast to the previous two analyses, the Template-Based Analysis relies primarily on three-dimensional, rather than two-dimensional source reconstruction, as described later in this document.

The three analyses presented in this paper search for neutrino events in Testbed data within the period from 
January 2011 to December 2012.
The Template-Based Analysis searches among all triggered events in the Testbed only from March to August of 2011.

All three analyses use a blinding technique to minimize individual bias determining analysis criteria.
Of the full data set, all software triggered events, all calibration pulser events, and only one out of every ten non-calibration pulser, RF triggered events are available for preliminary analysis and determining cut parameters.
Once the cuts are finalized, they are applied to the remaining 90$\%$ of the data.
The limits presented in the Results section represent the comparable results of the Interferometric Map Analysis and the Coherently Summed Waveform Analysis.

\subsection { Interferometric Map Analysis with Depth Dependent Ray Tracing }
\label{sec:IMA}

The first of the three analyses reconstructs events using an interferometric map technique.
For this analysis, we consider RF triggered events from  January 8$^{\rm{th}}$, 2011 to December 31$^{\rm{st}}$, 2012 and use a set of optimized cuts using AraSim calibrated against Testbed data to eliminate background events from our final sample.  
This analysis is performed in two stages.  
Stage 1 was a complete analysis on a limited data set that had been processed at an early period of data processing. 
A complete analysis is carried on data from from February-June of 2012 only, optimizing cuts on the 10\% set before opening the box on that time period alone.  
Then, in Stage 2 the analysis is expanded to the remainder of time in the two year period once 
more processed data became available.
In Stage 2,  the cuts were re-optimized on the 10\% set for the two year period but excluding February-June 2012 which had already been analyzed.  Table~\ref{tab:events} summarizes the number of events passing each cut.

Here we begin a description of all of the cuts applied in the Interferometric Map Analysis.
In both stages of the analysis, first we apply a few initial Event Quality Cuts to reject anomalous electronics behavior.
The period from  April 6$^{\rm{th}}$ to  May 12$^{\rm{th}}$ 2012 was characterized by consistent instability in the digitization electronics and thus this period is excluded as well.
We also remove events with corrupted waveforms, which comprise $\sim1\%$ of the entire data set.
We also reject an event if, in two or more channels, the power in frequencies below the high-pass-filter cut-off frequency of 150~MHz is greater than 10$\%$ of the waveform's power. 
This cut is designed to eliminate specific electronics errors that are otherwise difficult to identify. 
The percentage of simulated neutrino events rejected by this cut is less than 1\%.

The Interferometric Map Analysis includes a set of cuts that reject events independently of the signal strength.
This set is called the ``Effective Livetime Cuts" and consists of three cuts.
The first, the ``Calibration Pulser Timing Cut", rejects events that triggered within 80~ms centered around the beginning of the GPS second, which is when the calibration pulser signal is expected to arrive.
This 80~ms window is conservatively chosen for the Testbed analysis due to the rarely occurring electronic jitter in the system that makes the calibration pulser signal arrive later or earlier than its expected arrival time.
The Calibration Pulser Timing Cut reduces the efficiency by $\sim 8\%$.
The second cut is called ``IceCube Drilling Season Cut".
During the 2011-2012 data-taking period, IceCube was still actively drilling, and based on the time log of IceCube drilling, we conservatively reject the events occurring between October 22nd and February 16th each year.
The IceCube Drilling Season Cut rejects $\sim 31.4\%$ of the events.
The last cut is the ``Good Baseline Cut".
The Continuous Waveform (CW) Cut (described later) also requires the calculation of an average spectrum for each run to serve as a ``baseline" for the cut.
Before a baseline is determined to be an acceptable representation of the average background for a given run, we examine the characteristics of the run overall to determine if it is contaminated by a large number of CW events.
To do this, we calculate the maximum correlation between waveforms of neighboring events. If a significant number of CW events are found, they will be highly correlated with each other.
We do not use baselines in which more than $10\%$ of neighboring events are well correlated (contain a correlation between any two of the same antennas between events $>0.2$). 
In this case, we use an acceptable baseline from a nearby time period instead.
If no such baseline can be found, the Good Baseline Cut rejects the entire run, as it is likely to contain significant CW contamination.
This requirement reduces the efficiency by $\sim 15\%$.
Overall, the Effective Livetime Cuts reject $\sim 46\%$ of events independently of the signal strength.

We attempt a directional reconstruction for each event using the relative timing information and maximizing a summed cross-correlation over a set of hypothesized source positions.
We perform a cross-correlation on the waveforms from each pair of antennas of the same polarization.
This cross-correlation function measures how similar the two waveforms are with a given offset in time and is similar to the one described in~\cite{Romero-Wolf:2014pua}.
For each pair of antennas, we calculate the expected delays between the signal arrival times as a function of the position of a putative source relative to the center of the station.  The center of the station is located at the mean of the antenna positions $\sim20$~m deep in the ice.  
These signal arrival times account for the depth-dependent index of refraction in the firn layer and the abrupt change at the ice-air interface. If there are two ray trace solutions, only the direct one is considered.
For a given source position, the cross-correlation
value for an antenna pair is given by:\\
\begin{center}
\begin{equation}
C = \frac{ \sum\limits_{i = 1}^{N_{\rm{bins}}}{V_{1,i} \cdot V_{2,i+t} } }{ \sqrt{\sum\limits_{i=1}^{N_{\rm{bins}}}{V^2_{1,i}}} \cdot \sqrt{\sum\limits_{i=1}^{N_{\rm{bins}}}{V^2_{2,i+t}} } }
\end{equation}
\end{center}
where $V_{1,i}$ is the voltage in the $i^{\rm{th}}$ bin at the first antenna in the pair and $V_{2,i+t}$ is the voltage in the $(i+t)^{\rm{th}}$ bin at the second where $t$ is the number of bins corresponding to the expected time delay between the antennas for a signal from the putative source position
accounting for time dependences due to ray tracing.
Then, for each source position, the correlation values for each pair of antennas of the same polarization are weighted by the inverse of the integrated power of the overlap between the waveforms and Hilbert-transformed before being summed together to make the summed cross-correlation.

\begin{figure}
\centering
\hspace{-0.5cm}\includegraphics[width=0.48\textwidth]{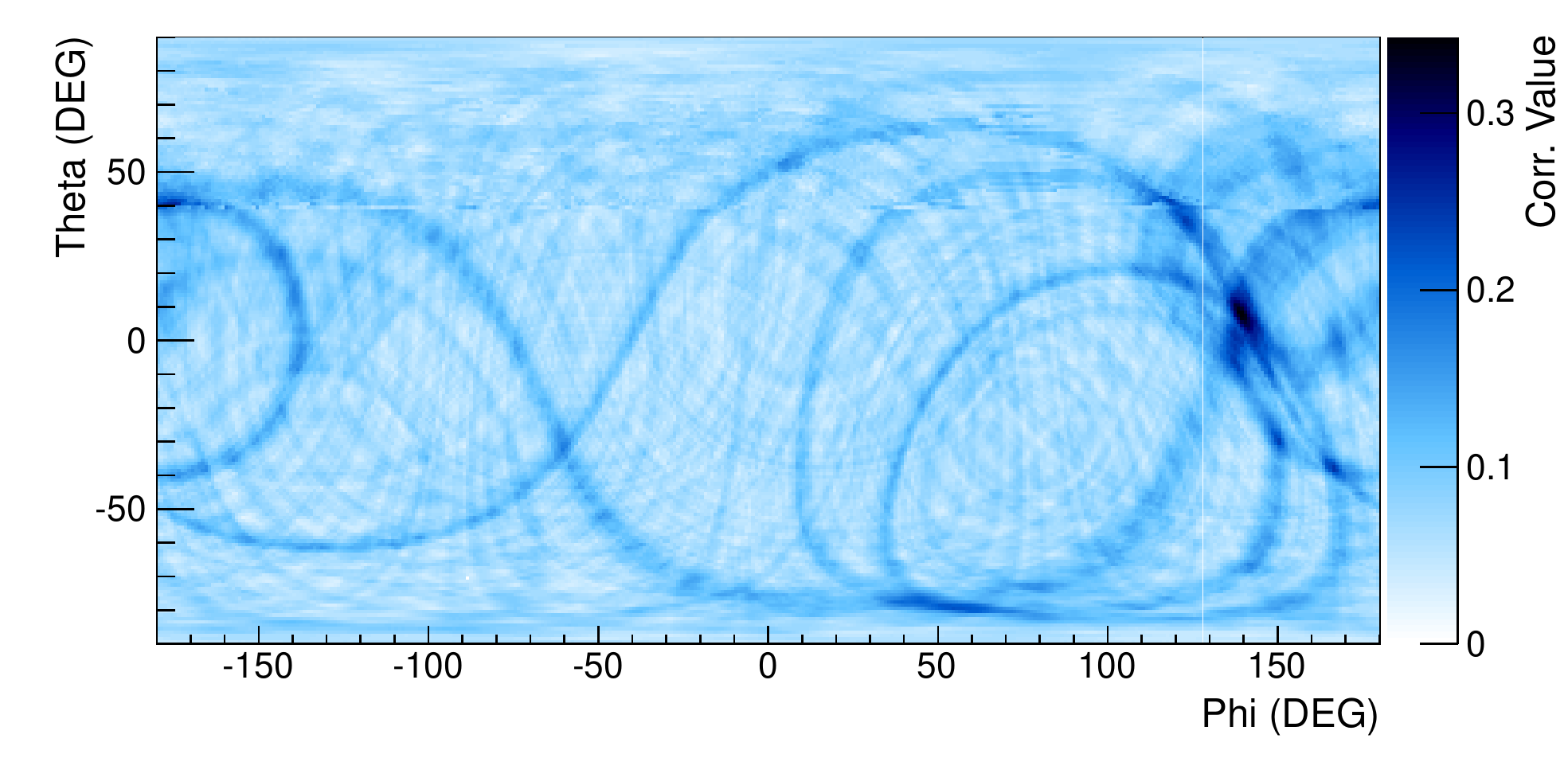}
\caption{\label{fig:interferometric_map} An example of a interferometric map used for reconstruction. This event is a calibration pulser and reconstructs well to the appropriate location in this 30~m inteferometric map where the correlation value is high (dark blue). }
\end{figure}

Summed cross-correlations for each polarization for each event are summarized in maps with 1$^\circ \times$1$^\circ$ bins in zenith and azimuth
for source distances of 3~km and 30~m only (see Fig. \ref{fig:interferometric_map}). 
 We use the 30~m map to determine if the event is a calibration pulser signal that has not been properly flagged, and the 3~km map to determine the reconstruction direction of sources  hypothesized to be from far away, such as neutrino events.
For each 1$^\circ \times$1$^\circ$ bin, we sample the correlation function for each pair of antennas at the 
delay expected for that source direction and distance, and form the summed cross-correlation that is
entered on the map.
We define the reconstruction direction to be the location of the peak in the correlation map.
Based on the calibration pulser events, our pointing resolution on the RF direction is $\sim1^\circ$.

Based on the reconstruction map generated by the interferometric reconstruction technique just described, 
we decide whether the map is of good quality in terms of pointing directionality.
When the signal is coming from one specific location and generates a consistent pattern of waveforms across multiple channels, the reconstruction map will point back to the direction of the source location.
A set of Reconstruction Quality Cuts
ensure that the event can be characterized by a single well-defined pointing direction that does not have an overly broad spot size on the map.  Thermal noise events will not exhibit this strong well-defined peak in an interferometric map.
The Reconstruction Quality Cuts also require that any strong correlation is not found only in a single bin on the map with no other comparable sized correlation values nearby. 
This would not be consistent with the antenna and electronics responses of the detector.  

The Reconstruction Quality Cuts are based on an area  surrounding the peak correlation where the correlation remains high, $\rm{A_{peak}}$,
and the total
area on the map showing high correlations, $\rm{A_{total}}$.
We first find the $85\%$ contour surrounding the point of peak correlation and the area of that contour in square degrees is $\rm{A_{peak}}$. 
The total area on the map that shows a correlation higher than $85\%$ of the maximum correlation value is called $\rm{A_{total}}$.
The first Reconstruction Quality Cut condition requires the size of $\rm{A_{peak}}$ to be greater than 1~deg$^2$ and less than 50~deg$^2$.
The minimum of that range is the area of a single bin on the map, due to individual time bins in the waveforms.  
The 50~deg$^2$ was chosen because in a distribution of 
$\rm{A_{peak}}$ from cal pulser events, it was $\sim2~\sigma$ away from the mean of the distribution at $\sim30$~deg$^2$.
The choice of the 85\% level for the contour was somewhat arbitrary.  A different choice would have led to a different maximum allowed
$\rm{A_{peak}}$.
Note that the area of the $85\%$ contour around the peak is not the same as the resolution of the reconstruction.
Instead, the area of the contour is related to the width of the readout impulse.
The second condition for the Reconstruction Quality Cut requires the ratio between $\rm{A_{total}}$ and $\rm{A_{peak}}$ to be less than 1.5.   This means that only one reconstruction direction dominates the map.

Each event is separated into VPol and HPol channels. 
A VPol or HPol channel is required to pass these two Reconstruction Quality Cut conditions in the 3000~m maps.
For the purpose of tabulating results, the rest of the cuts are applied to VPol and/or HPol channels separately after the Reconstruction Quality Cuts.  
The cuts applied to each polarization are the same and for any event, one or both channels may pass the cuts.

\begin{figure}[t]
  \centering
  \includegraphics[width=0.5\textwidth]{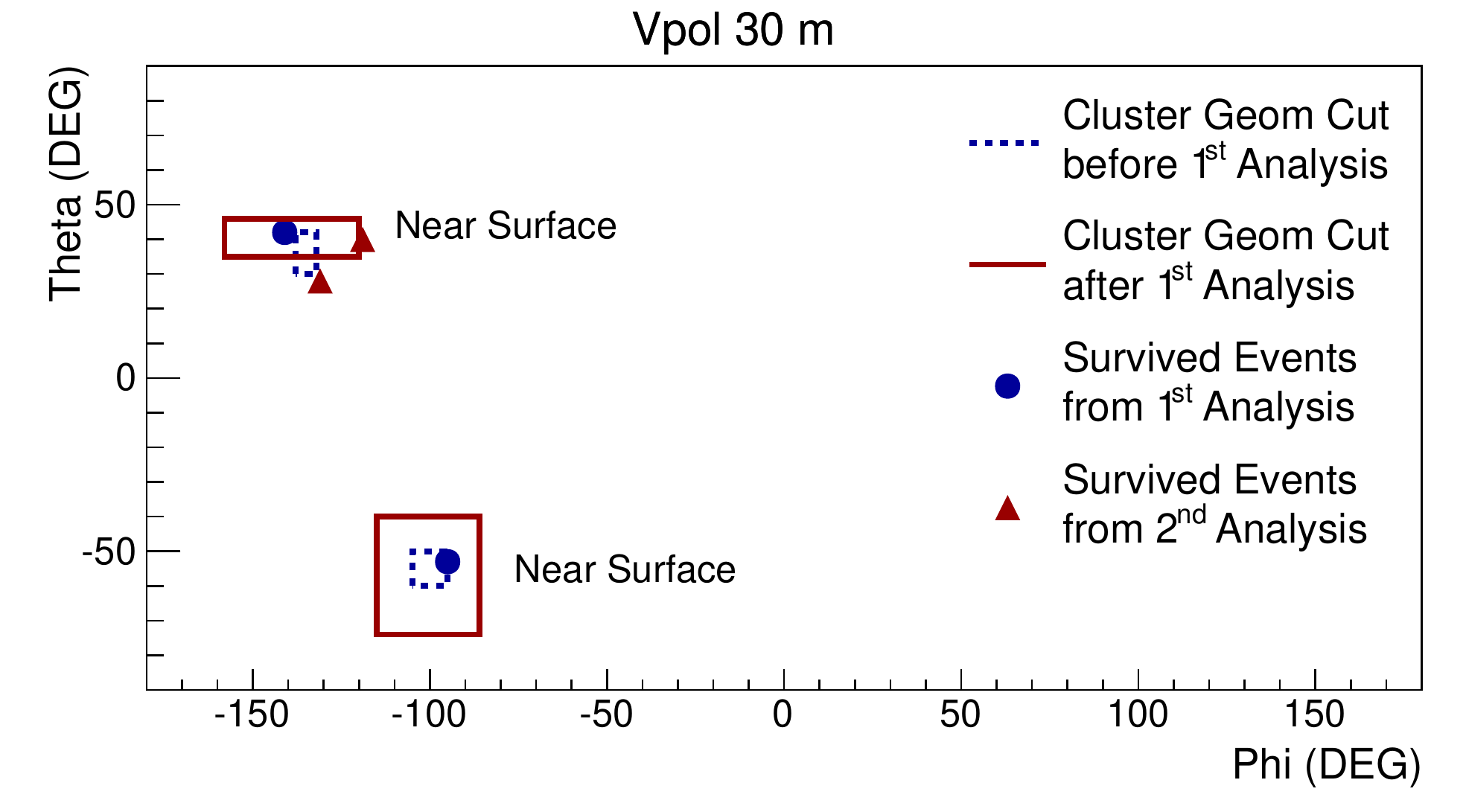}
  \includegraphics[width=0.5\textwidth]{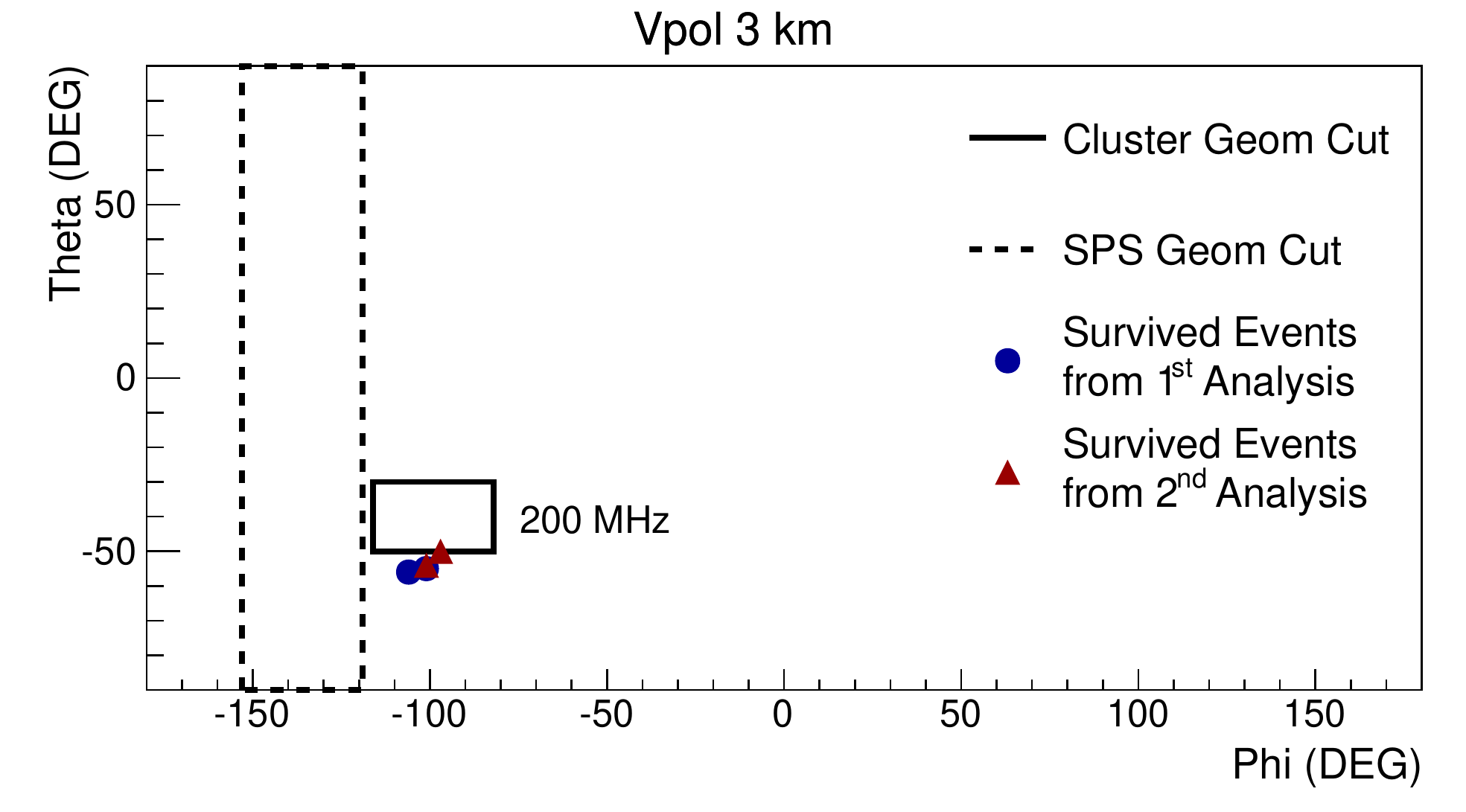}
  \caption{The reconstructed directions of the events that passed both Stage~1 and Stage~2 of the analysis in the 30~m (upper) and 3~km (lower) maps. Events that passed the unaltered cuts in Stage~1 are shown in blue and those that passed the Stage~2 cuts are shown in red. The initial Geometric Cut regions (dashed blue line) were adjusted after Stage~1 (solid red lines) based on 
  a  Gaussian fit to the background event distribution with a limited set of cuts applied.}
  \label{fig:modified_geom_cuts}
 \end{figure}

A set of Geometric Cuts reject events that reconstruct to locations
where background due to anthropogenic noise is expected to be high, either where there is known human activity or where
signals reconstruct to the same location repeatedly.
See Fig.~\ref{fig:modified_geom_cuts}. 
The reconstructed directions used for this cut are derived using the interferometric reconstruction technique described above. 
Events that reconstruct to South Pole Station (SPS) are rejected. 
This area covers a region of -153$^\circ$ and -119$^\circ$ in azimuthal angle. 
Events that reconstruct to within a box in zenith and azimuth 
centered around the location of a calibration pulser are also rejected (see below).
Additionally, we reject regions where
 where multiple events reconstruct after several other cuts have been applied. 
This removes signals originating from unknown but repeating sources.  Neutrino events are not expected to originate from the same position repeatedly. The events targeted by these Geometric Cuts, other than the calibration pulser events, may all be coming from sources at SPS as one of the best reconstructing directions is generally pointing toward SPS.

Three repeating locations were identified in the Interferometric Map Analysis, 
two  in the VPol 30~m map, and another on the VPol 3~km map.
The events reconstructing to a repeating location in the Vpol 30~m map are characterized as ``near surface" events as they appear to come from a point near the surface of the ice at $\theta \approx +40^{\circ}$ relative to the station center.
The two locations in the VPol 30~m map reject the same type of event where the best reconstructed location can be either of two locations depending on the strength of the signal.

The first of these two reconstruction locations for ``near surface'' events is 
centered at a zenith angle of 40$^\circ$ and an azimuth of 140$^\circ$.  We reject any events whose VPol 30~m reconstruction points within  a box that is 10$^\circ$ in zenith and 40$^\circ$ wide in azimuth and centered on that location.
The second of the two reconstruction locations for ``near surface'' events is centered at a zenith angle of -57$^\circ$ and an azimuth of -100$^\circ$.  We reject any events whose VPol 30~m reconstruction points within a 30$^\circ \times$ 30$^\circ$  box in zenith and azimuth surrounding that point.

One repeating location was identified in the VPol 3~km map and these events are characterized by an excess of power in a $\sim 50$~MHz band around 200~MHz, and as such are labeled ``200~MHz events."
This ``200~MHz" repeating region from the VPol 3000~m map is centered at a zenith angle of -40$^\circ$ and an azimuth of -99$^\circ$ with the rejection region being 20$^\circ$ wide in azimuth and 34$^\circ$ high in zenith.
As with the ``near surface" events, there is a secondary reconstruction point but it is located within the SPS reconstruction region and thus events that reconstruct there are already rejected. 
These events from repeating locations are expected to be more effectively rejected by other means after improving the reconstruction method in the next analysis.

\begin{figure*}[!ht]
\centering
\subfloat[]{\includegraphics[width=0.45\textwidth]{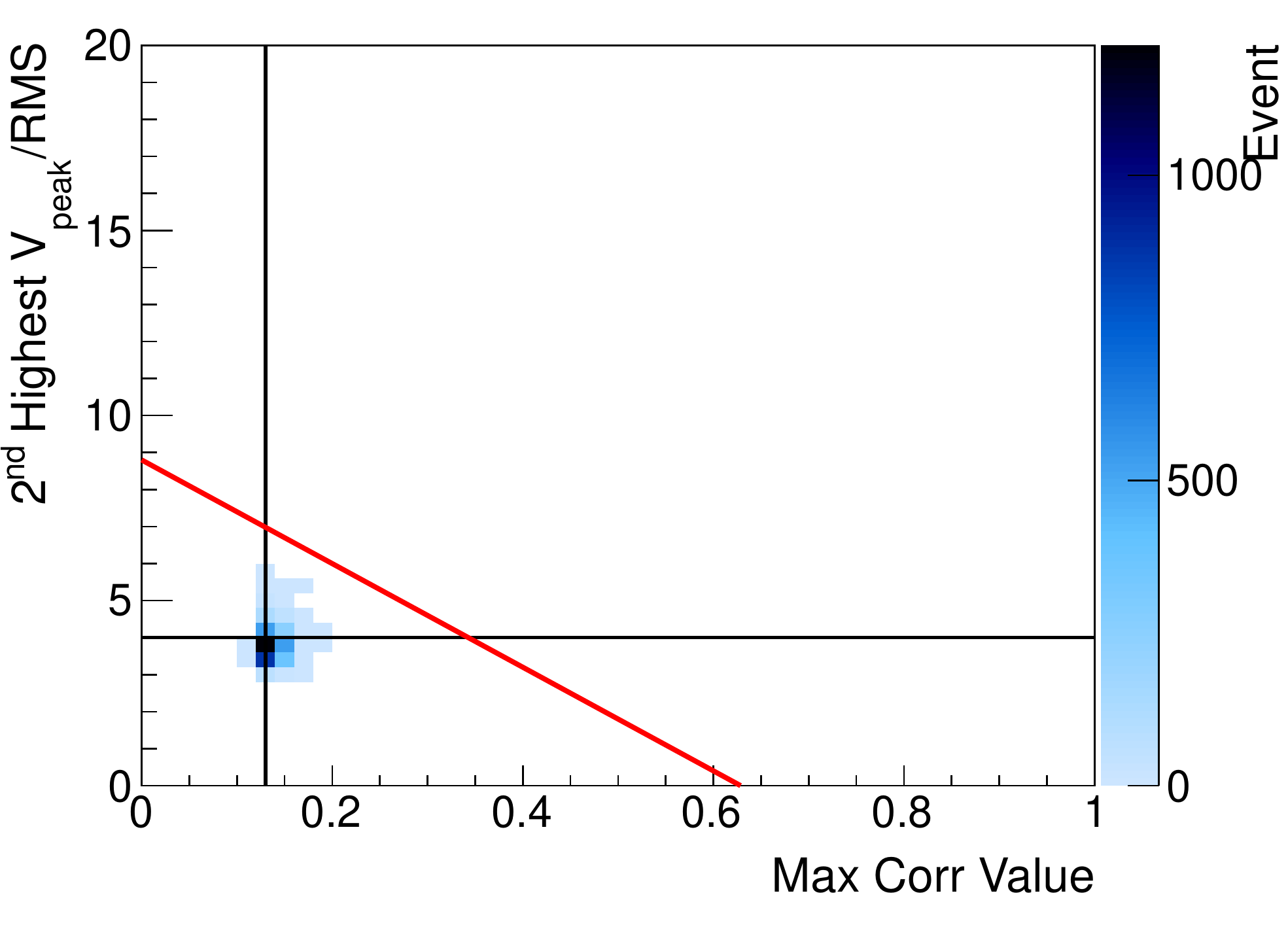}} \hfill
  \subfloat[]{\includegraphics[width=0.45\textwidth]{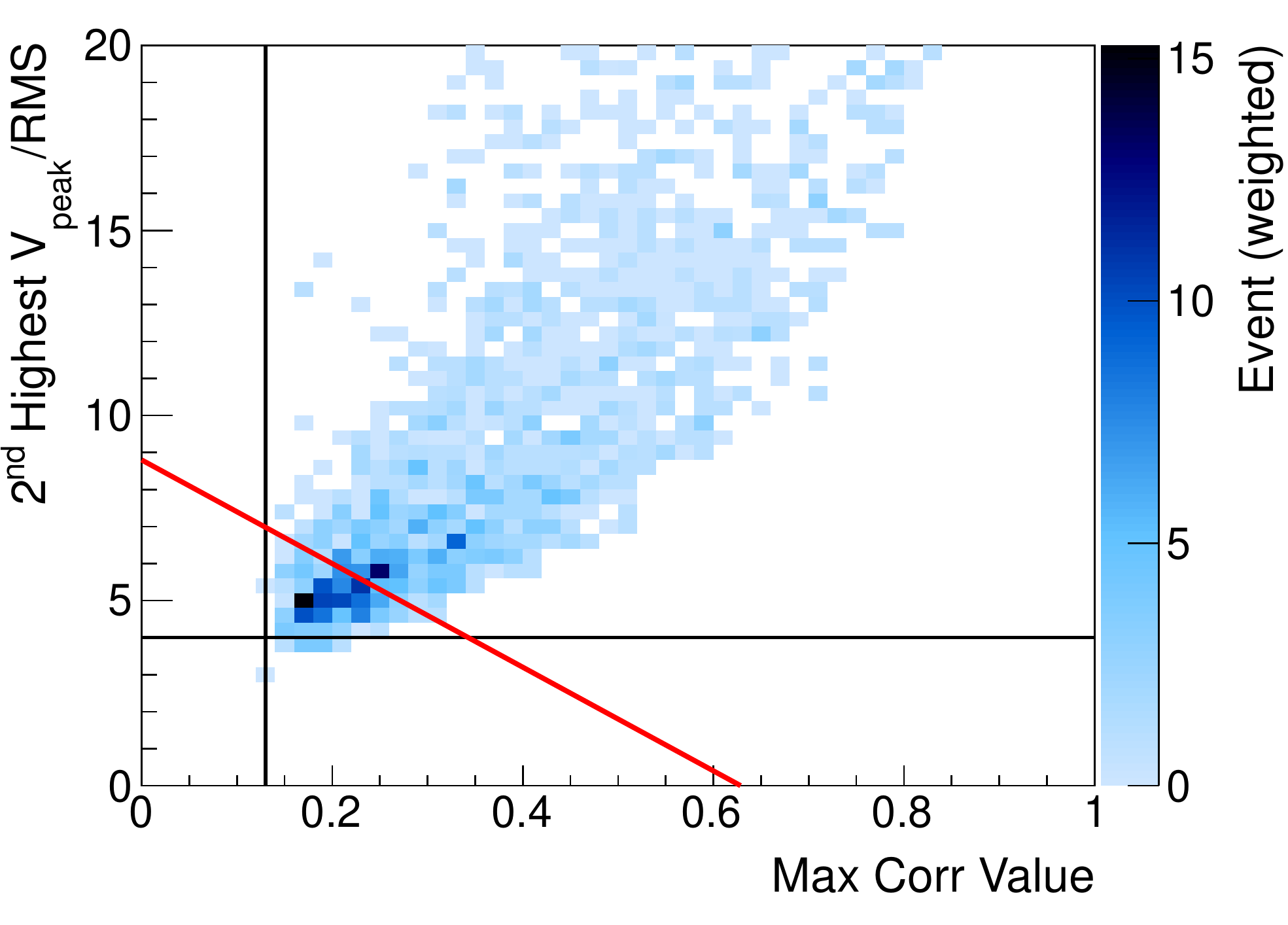}}
\caption{\label{fig:RCut} The distribution of $2^{\rm{nd}}$ highest $\rm{V_{peak}/RMS}$ and correlation values for the vertical polarization channel for (a) the 10\% examination data set and (b) events simulated at $10^{18}$ eV. Both plots show only events that have survived all other cuts. The red line shows the selected cut parameter and thus all events above this line survive the cuts and those below are removed. For the data (a), no events fall above the cut line. For the simulated events (b), there is a sizable percentage of events that lie above the cut line and thus survive the analysis. These simulated events extend to a range of higher correlation and $\rm{V_{peak}/RMS}$ values with a slight bias towards lower correlation values as events may misreconstruct.}
\end{figure*}

\begin{figure*}[!ht]
\centering
\subfloat[]{\includegraphics[width=0.45\textwidth]{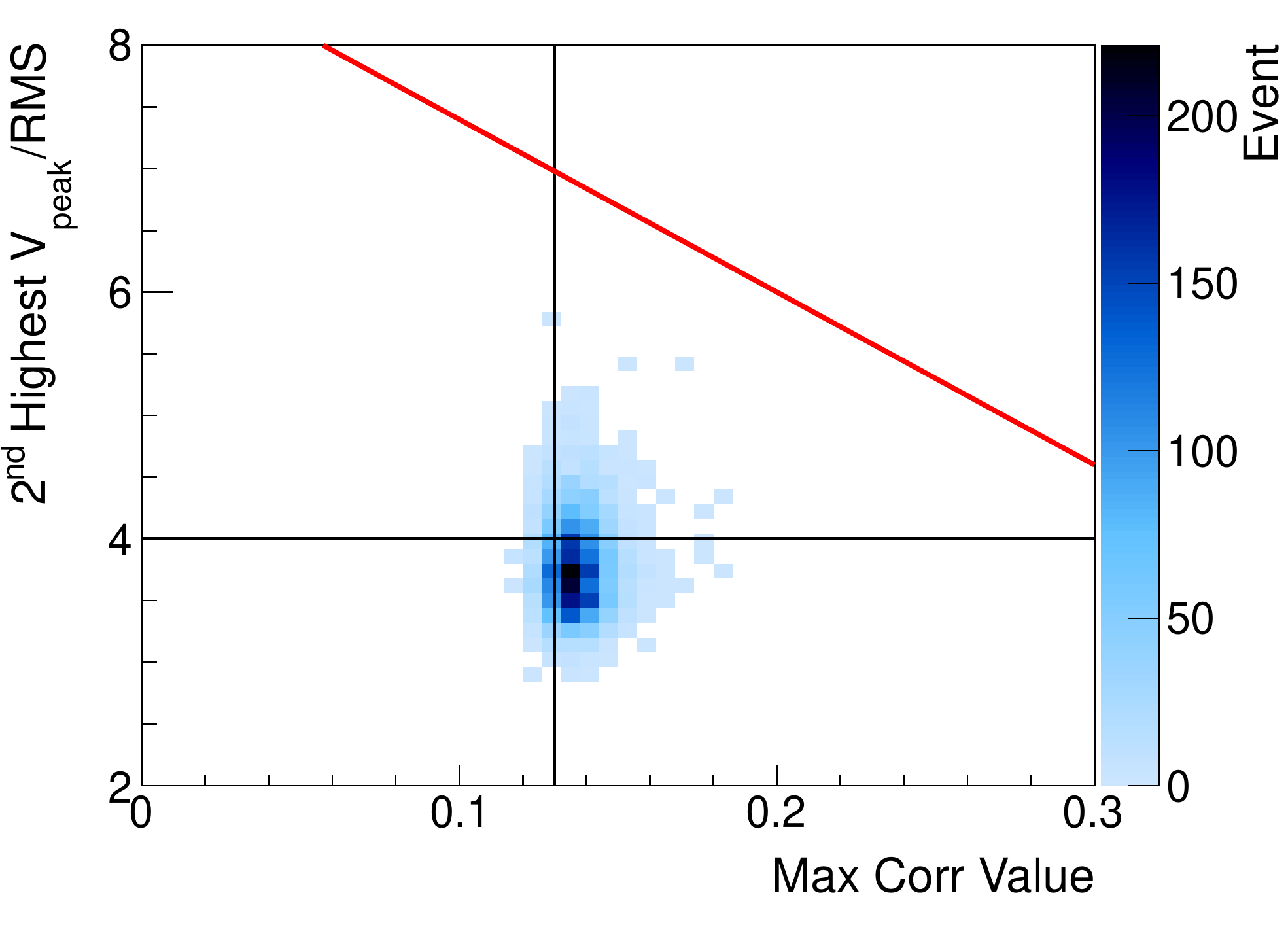}} \hfill
  \subfloat[]{\includegraphics[width=0.45\textwidth]{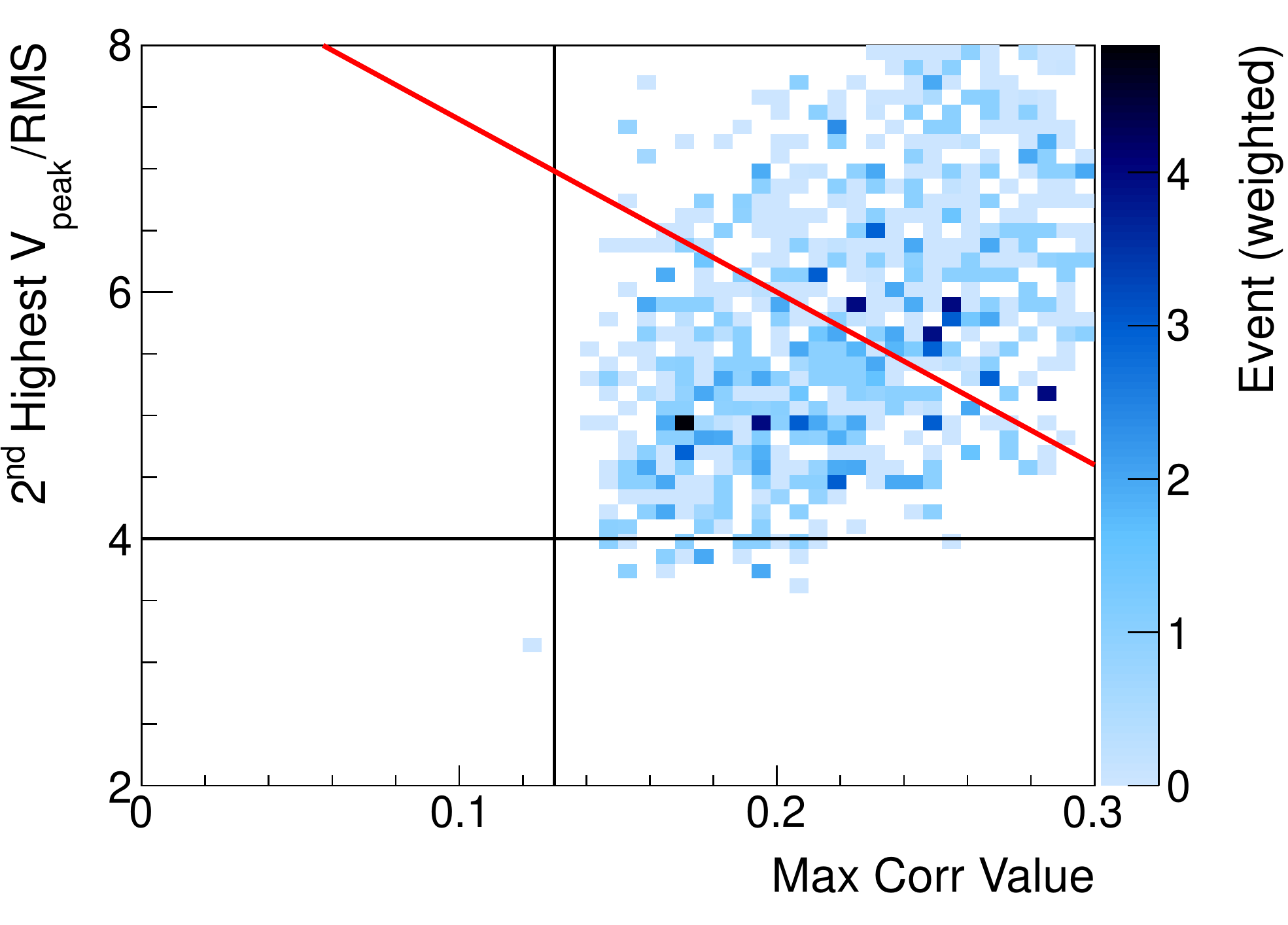}}
\caption{\label{fig:RCut_zoomed} These plots are zoomed in versions of the plots above (Fig. \ref{fig:RCut}). In the data (a), the events are dominated by thermal noise and thus concentrate around specific low correlation and $\rm{V_{peak}/RMS}$ values of 0.135 and 3.7, respectively. The simulated events are dominated by the simulated signal and thus do not tend to cluster around a particular value for the correlation and signal strength.}
\end{figure*}

A Saturation Cut rejects events when the saturation of the amplifier induces distortion of the waveform.
When the signal strength is strong enough to saturate the amplifier and change the linearity of the amplification factor, we may have misleading reconstructed information from the event and thus we want to remove it from consideration.
As the maximum dynamic range of the amplifiers of the borehole antennas is $\sim\pm$1~V, we set the saturation point of the output voltage to $\pm995~\rm{mV}$ and when two or more channels' waveforms have maximum voltage values that exceed the saturation point, we reject the event.

The Gradient Cut is a pattern recognition cut to reject a specific type of background event which has a strong gradient in signal strength across the Testbed in one direction.
The Geometric Cut was not sufficient to reject this specific type of background event because 
the Geometric Cuts are effective for sources located at $\sim30~\rm{m}$ or $>1~\rm{km}$ distance from the Testbed, while only
a closer source can give us such a large gradient in signal strength across the station.
We first check whether there is a gradient in signal strength in the direction that matches that of these background events.  One of the two VPol channels facing SPS has the strongest $\rm{V_{peak}/RMS}$ value for these events while one of other two VPol channels has the weakest signal strength.
When the pattern matches, we calculate the gradient value, G, given by
\begin{equation}
G = \frac{ |V_{\rm{max}} - V_{\rm{min}}| }{ \sqrt{ V^2_{\rm{RMS,max}} + V^2_{\rm{RMS,min}} } }~~~~~~~~,
\end{equation}
where $V_{\rm{max}}$ is the peak voltage of the channel with the highest $\rm{V_{peak}/RMS}$, $V_{\rm{min}}$ is the peak voltage of the channel with the lowest $\rm{V_{peak}/RMS}$, and $V_{\rm{RMS,max}}$ and $V_{\rm{RMS,min}}$ are the RMS voltages of those same channels respectively.
If the gradient value is greater than $3.0$ we reject the event.

The Delay Difference Cut ensures that the reconstruction direction derived from all the borehole antennas of the same polarization is
consistent with the delay observed between the signals in the two antennas with the strongest signals. 
This cut determines whether the reconstructed direction for the event corresponds to the timing difference between the highest peak voltages in the waveforms. 
In the case of an impulsive signal like a neutrino event, we expect this correlation to exist whereas a thermal noise event in general should not exhibit this behavior.
We calculate the  time delay $\Delta t _{\rm{1,2, peak}}$ between the peak voltages $\rm{V_{peak}/RMS}$ in the two
 channels with the highest peak voltages.
We also find the time delay that would be expected between those two channels based on the direction of reconstruction, $\Delta t _{\rm{1,2, reco}}$. 
We then find the difference between these two values, $\Delta \rm{T_{delay}} = \Delta t _{\rm{1,2, peak}} - \Delta t _{\rm{1,2, reco}}$.
 If $|\Delta \rm{T_{delay}}| > 20~\rm{ns}$, we reject the event.

The In-Ice Cut rejects the events that reconstruct to directions above horizontal as viewed by the Testbed.
This cut is made because we are searching for neutrino events that are coming from the ice.

A CW Cut rejects the events that are contaminated with narrowband anthropogenic noise.
This cut rejects events that show a narrowband peak above an expected noise spectrum or baseline as described earlier.
Then for each channel, we compare the individual event's waveforms against the baseline.
A frequency bin is flagged as containing CW if, that bin in one channel exceeds 6.5~dB above the baseline, and two other channels also have a bin within 5~MHz of the first that exceeds the threshold.
In order to maintain a high efficiency for neutrinos, we require that this excess is narrowband before rejecting the event.
We define a signal to be narrowband if less than $50\%$ of frequency bins in a 40~MHz band around the peak are above 6.0~dB above the baseline.

As a last cut, a Peak/Correlation Cut is applied.
Since we expect impulsive events to exhibit a correlation between the $\rm{V_{peak}/RMS}$ values from the waveforms and maximum correlation value from the reconstruction map, we designed a cut using 
these two values, as in~\cite{Gorham:2010kv,Gorham:2010xy}.
CW-like events tend have high correlation values but low $\rm{V_{peak}/RMS}$ values.
Conversely, thermal noise events may fluctuate to high $\rm{V_{peak}/RMS}$ values but not correlate well in any particular direction.

The Peak/Correlation Cut is based on a 2-dimensional scatter plot that 
has $2^{\rm{nd}}$ highest $\rm{V_{peak}/RMS}$ on the vertical axis and a maximum correlation value on the horizontal axis for the corresponding polarization (see Fig.~\ref{fig:RCut} and Fig.~\ref{fig:RCut_zoomed}).
We choose the $2^{\rm{nd}}$ highest $\rm{V_{peak}/RMS}$ value from the waveforms in order to ensure that the value represents the signal strength in at least two channels and not a random fluctuation from thermal noise.
First, we set constant cuts at  $2^{\rm{nd}}$~highest $\rm{V_{peak}/RMS} > 4.0$ and maximum correlation value $ > 0.13$.
We use the $2^{\rm{nd}}$ highest $\rm{V_{peak}/RMS}$ instead of the highest so that two channels exceed our threshold.
After this, we define a cut as a line on the plot of $\rm{V_{peak}/RMS}$ vs. maximum correlation as shown in the figures 
(red line in Fig.~\ref{fig:RCut}).
Events located above this line will pass the cut.

\begin{figure}
\centering
\includegraphics[width=3.0in]{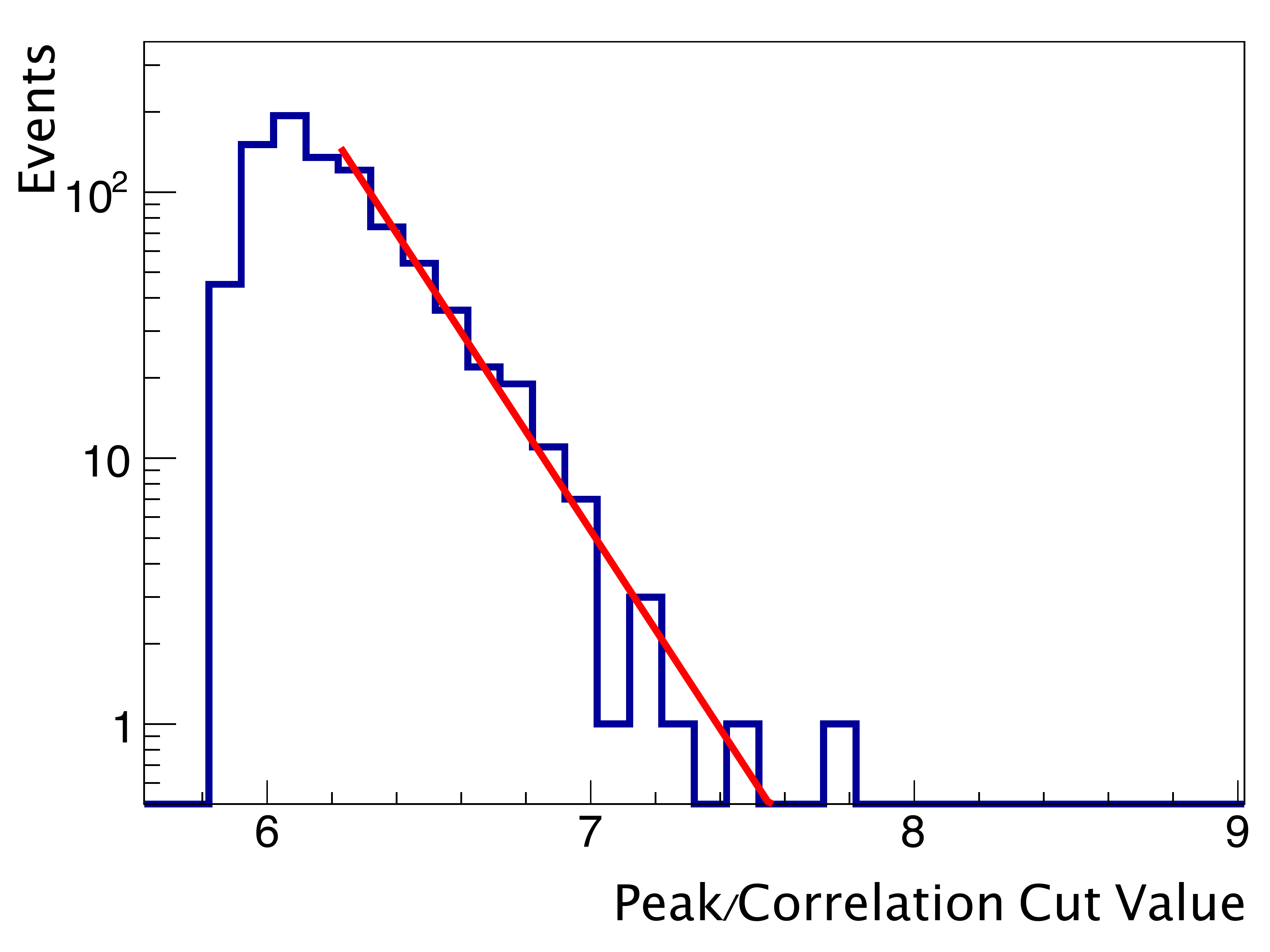}
\caption{\label{fig:PeakCorr_bestfit} The differential distribution of events that pass the Peak/Correlation Cut using the optimal slope shown in Fig.~\ref{fig:RCut}.  The horizontal axis, Peak/Correlation Cut Value, is a measure of the vertical offset of the red line in Fig.~\ref{fig:RCut}, and is the $2^{\rm{nd}}$ highest $\rm{V_{peak}/RMS}$ peak where the red line intersects the Max Corr Value=0 axis.  This distribution is fitted against an exponential function which is used to extrapolate to the number of events expected to pass the cut.}
\end{figure}

We chose a slope that gives a reasonable p-value on an exponential fit to the differential number of events passing the cut as a function of the position of the vertical line (see Fig.~\ref{fig:PeakCorr_bestfit}).
We tested two different slopes, -14 and -9, which each give a reasonable p-value (of order 0.1-1), and chose the one that gave the best expected limit on the Kotera maximal model~\cite{Kotera:2010yn} (Faranoff-Riley type II strong source evolution~\cite{Wall:2004tg} with a pure proton composition).
We choose a slope of -14, which gives a p-value of 0.235.
We defined the Peak/Correlation Cut Value as the vertical offset of the line, defined as the intersection between the slope and the Max Corr Value=0 axis in Fig.~\ref{fig:RCut}.
We then optimized the slope and the Peak/Correlation Cut Value against the Kotera maximal model.
The optimal slope and Peak/Correlation Cut Value were found to be -14 and 8.8, respectively and is shown in Fig.~\ref{fig:RCut} and Fig.~\ref{fig:RCut_zoomed}.

To estimate the background, we use the 10$\%$ data set and fit  the differential number of events rejected by the Peak/Correlation Cut as a function of 
Peak/Correlation Cut Value, shown in Fig.~\ref{fig:PeakCorr_bestfit}. 
We use the fit function $N_{\rm{diff}} = e^{a\cdot x + b}$ where $x$ is 
Peak/Correlation Cut Value,
$N_{\rm{diff}}$ is the differential number of events rejected by the Peak/Correlation Cut when 
Peak/Correlation Cut Value change by $dx$, 
and $a$ and $b$ are two fit parameters in the exponential function.
From the fit, we obtained 
$a= -4.29 \pm 0.26$ 
and $b=31.70 \pm 1.67$  where the deviation of each parameter is the one sigma error from likelihood fit result.
The optimal vertical offset gives us 0.03 estimated background events and 0.01 expected neutrino events from the Kotera maximal model in the 90\% data set in the Stage 2 analysis.

\begin{figure*}[!ht]
\centering
\subfloat[]{\includegraphics[width=0.5\textwidth]{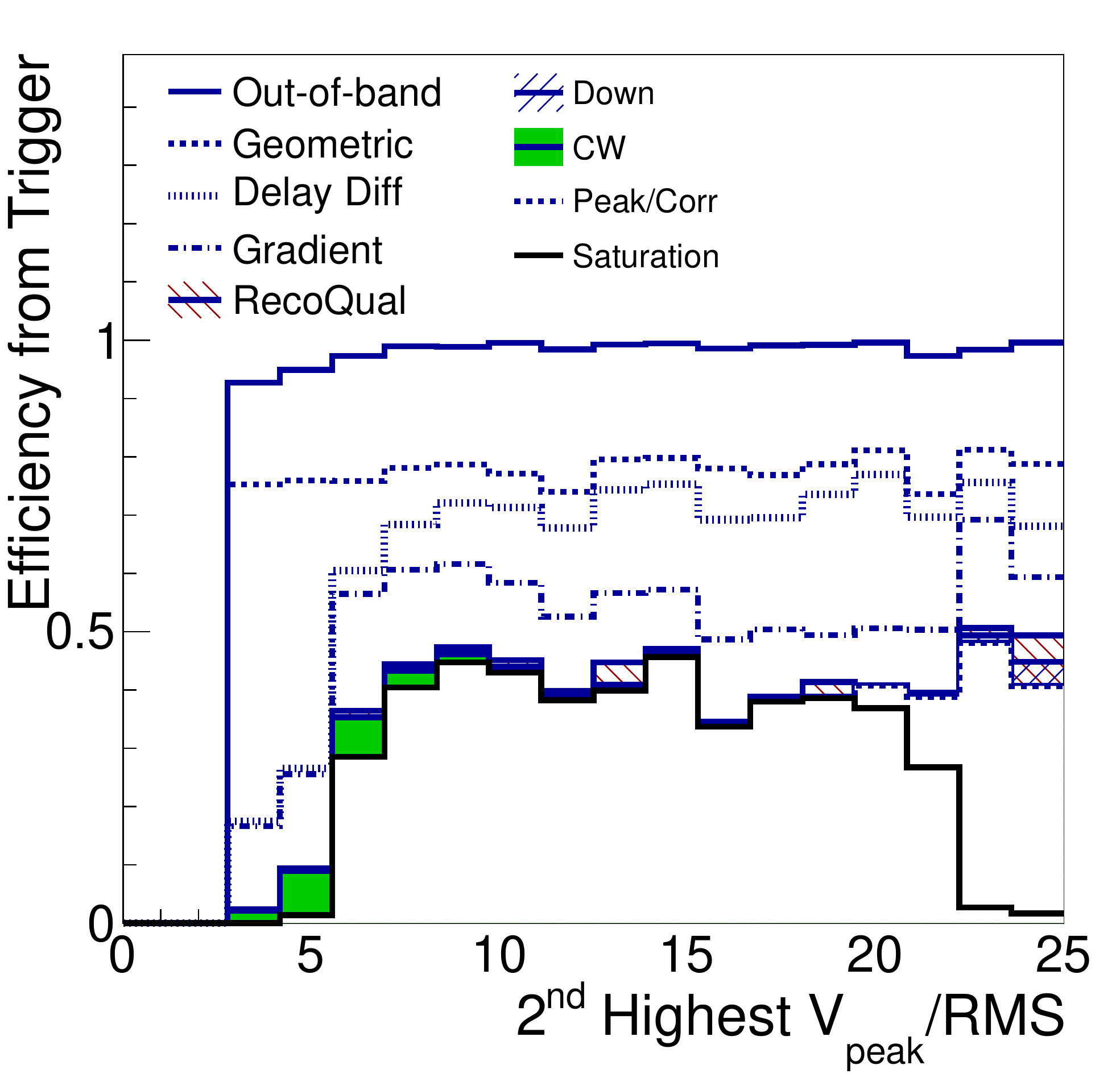}} \hfill
  \subfloat[]{\includegraphics[width=0.5\textwidth]{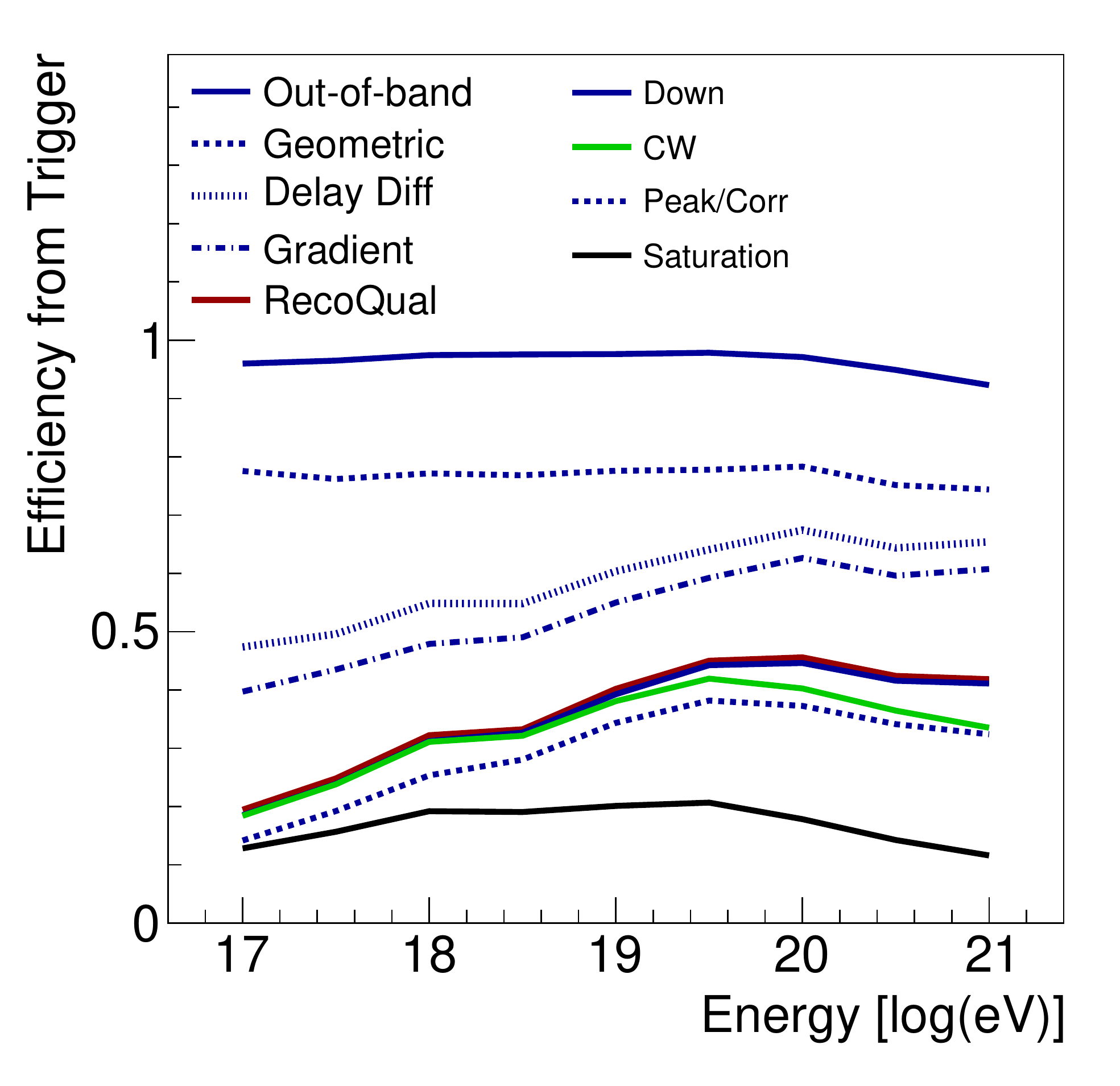}}
\caption{\label{fig:cut_efficiencies} 
The cumulative efficiency of each cut on triggered events from a simulated data as a function of signal strength measured as the 2$^{\rm{nd}}$ highest $\rm{V_{peak}/RMS}$ (a) and neutrino energy (b).
Note that these plots do not include the effect of the Effective Livetime Cuts, which together contribute to a 46\% loss in analysis efficiency independent of signal strength.
}
\end{figure*}

Fig.~\ref{fig:cut_efficiencies} shows the cumulative efficiencies after each analysis cut for triggered events as modeled in AraSim for the Interferometric Map analysis.
For this plot, we use the 2$^{\rm{nd}}$ highest $\rm{V_{peak}/RMS}$ as in the Peak/Correlation Cut and neutrino energy on the horizontal axis for plot (a) and (b) respectively.
Cuts in Fig.~\ref{fig:cut_efficiencies} do not include the Effective Livetime Cuts since they do not depend on signal strength.
Since the Effective Livetime Cuts reject 46\% of events independent of signal strength, the total analysis efficiency is $\sim 20\%$ for signal strengths between 7 and 20 in 2$^{\rm{nd}}$ highest $\rm{V_{peak}/RMS}$ in Fig.~\ref{fig:cut_efficiencies} (a).
Similarly, the efficiency in Fig.~\ref{fig:cut_efficiencies} (b) plateaus from $10^{18}$ to $10^{19.5}$~eV, becoming $\sim 10\%$ after including the Effective Livetime Cuts.
Both Fig.~\ref{fig:cut_efficiencies} (a) and (b) show that the Saturation Cut rejects most events at signal strengths above 20 $\rm{V_{peak}/RMS}$ and neutrino energies above $10^{19.5}$~eV.
The shadowing effect due to ray tracing (Fig.~\ref{fig:RayTrace}) limits the detectable volume to nearby ice for the shallow Testbed and this means that at high neutrino energies a large fraction of detectable events will saturate the amplifier.
The efficiencies for high SNR are $\sim 40\%$ without the Effective Livetime Cuts and $\sim 20\%$ with the Effective Livetime Cuts.

In Stage~1 of the analysis, we had three events survive all cuts.
These three events were all known types of anthropogenic impulsive events, and one
was removed by altering the Gradient Cut and the other two by 
altering the Geometric Cuts.
In Stage~2 of the analysis (2011-2012), using these new Geometric Cut regions, two events survived. 
The four events that were rejected by the modified Geometric Cuts in the first and second stages can be seen in Fig.~\ref{fig:modified_geom_cuts}, along with the Geometric Cut regions.
The alterations to the Geometric Cut regions increase the total acceptance of the Geometric Cut  (which includes the south pole region) by less than $5\%$.
After these modifications, zero neutrino candidate events survived.  In future analyses, we plan to design cuts to reject these type of events by other means, with less reliance on the Geometric Cuts. 

In Stage 1, one of the three events that passed appears to be a ``200~MHz'' type event, and
 we had intended to reject those with the Gradient Cut and this cut was modified slightly to better match the pattern that it was trying to identify.
The original definition of the cut required that the highest $\rm{V_{peak}/RMS}$ value came from a VPol channel before the gradient condition was checked for the event.
Through what appears to be an aberrant single-bin fluctuation, this event had its highest $\rm{V_{peak}/RMS}$ in an HPol channel.
Since the requirement that the highest $\rm{V_{peak}/RMS}$ value be from a VPol channel is not a necessary condition for the pattern recognition, it was removed from the definition of the cut.
The adjusted cut just checks that the gradient among the VPol channels matches the pattern and using this adjusted definition, the event was then rejected.

The other two events that passed the Stage 1 analysis cuts corresponded to the ``Near Surface'' event type and appears to be one intended to be removed by the Geometric Cuts.
Initially the location and size of these regions were defined  by eye, but after Stage~1 of the analysis, they were adjusted in a more quantitative manner as described in the next paragraph.

The two altered ``Near Surface'' regions were defined after performing Gaussian fits to the azimuth and zenith distributions of the events with only the 
Event Quality, Reconstruction Quality Cut, Delay Difference and CW Cuts applied.
The edges of each cut region were defined by the criterion that one would expect a total of 0.02 background events to reconstruct outside the region based on this Gaussian fit with only these four cuts applied.
This modification increased the total size of the Geometric Cut area, including the SPS, Calibration Pulser, and Clustering Cuts, by $\sim14\%$. 
After these adjustments, all of the events were rejected.

In Stage~2, the two events that passed were again leaked anthropogenic impulsive events that were intended to be rejected by the ``200 MHz'' Geometric Cut and thus were removed by slightly expanding the Geometric Cut region in the 3~km map.

\subsection { Coherently Summed Waveform Analysis }
\label{sec:CSW}

The Coherently Summed Wave Analysis differs from the Interferometric Map Analysis in its
reconstruction method, continuous wave rejection and other cut parameters. 
The initial data quality cuts and trigger timing cuts (to reject calibration signals) are performed in a similar manner and will not be discussed here.

Although affecting only a small fraction of livetime and recorded events, CW sources are a significant background to any physics search. 
The concept of ARA relies on coherence between signals seen in multiple antennas, and continuous wave sources, such as communications signals, provide strong coherence. 
As these signals are a background to the analysis a CW Cut is implemented to reject events that have characteristics similar to CW signals. 
A probability is calculated on an event by event basis that the measured frequency content is thermal in nature. 
Events are then rejected when an excess is observed  across a narrow range of frequencies. 
The probability threshold and minimum width are tuned using a combination of events identified as containing a known CW source, calibration pulser events and simulated neutrino signals to avoid rejecting broadband signals.

\begin{figure*}[!ht]
\centering
\subfloat[]{\includegraphics[width=0.5\textwidth]{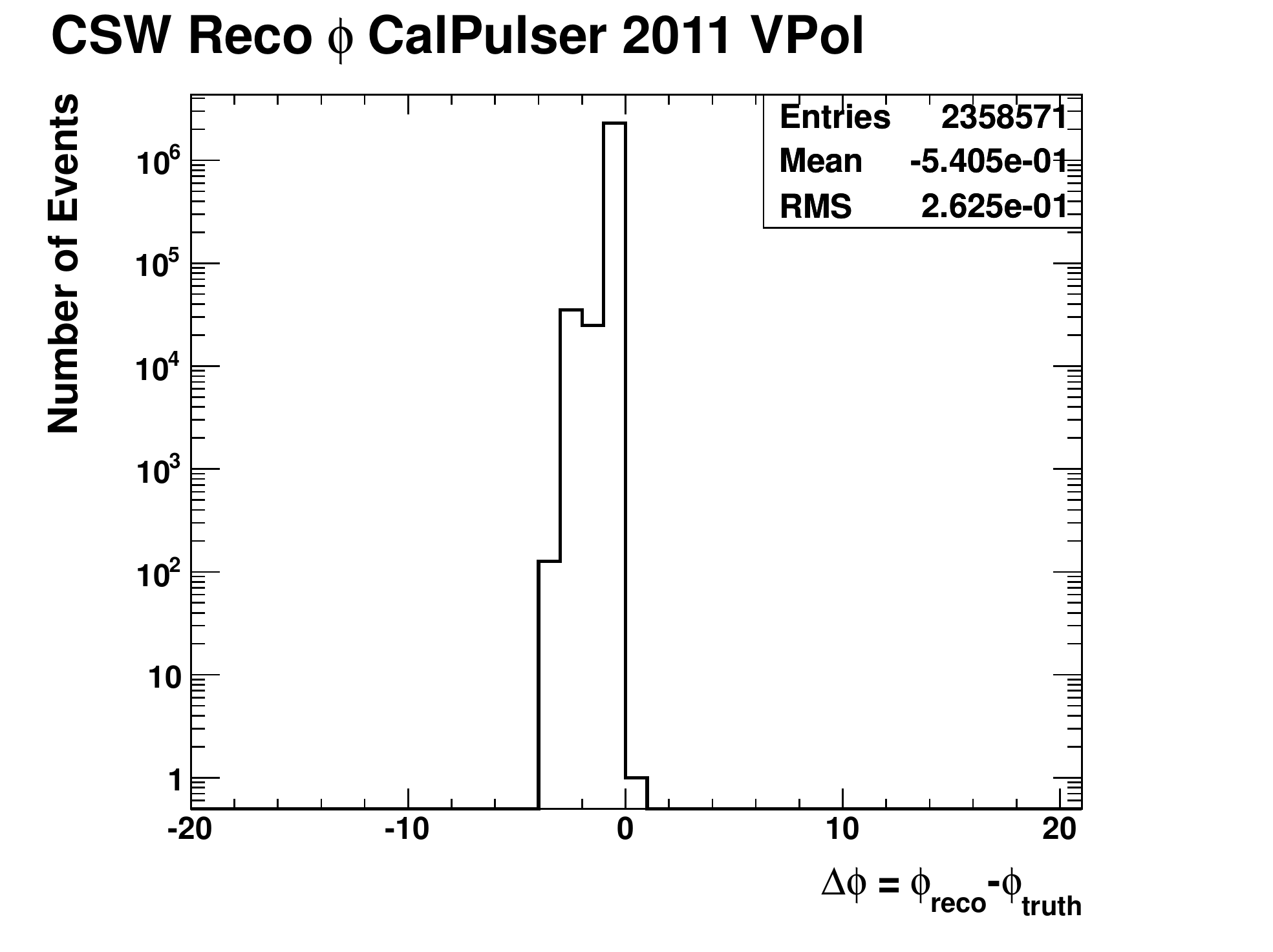}} 
  \subfloat[]{\includegraphics[width=0.5\textwidth]{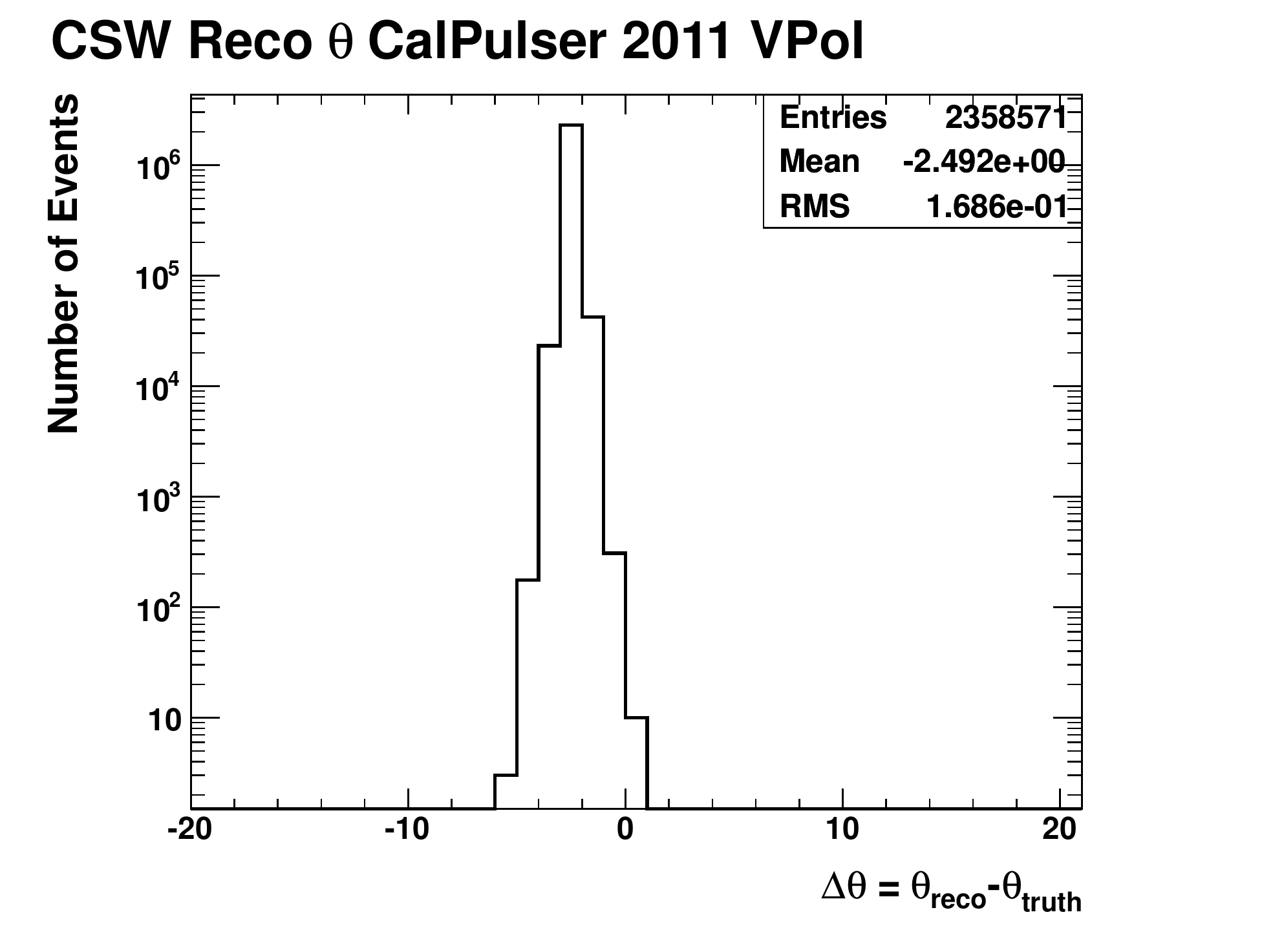}} \\
\caption{\label{fig:csw-reconstruction} Residuals for reconstructed source direction of an in ice TestBed calibration source. Shown in (a) is the azimuth ($\phi$) and in (b) the elevation ($\theta$) angle reconstructed minus the true location of the source.}
\end{figure*}

The reconstruction method is based upon calculating timing offsets between waveforms that maximize correlation. 
This is achieved by creating a coherently summed wave (CSW), where individual antenna waveforms are added, offset in time relative from one another. 
These offsets are computed using a simple algorithm that finds a CSW that is maximally correlated with the individual antenna waveforms, as measured by cross-correlation. The CSW is defined as:

\begin{equation}
  \mbox{CSW}(t) = \frac{1}{N}\sum_{i} \psi_{i}(t+\Delta t_{1,i})\\
\end{equation}
where  $N$ is the total number of antennas, $\psi_{i}$ is the time domain waveform for the $i^{th}$ antenna and $\Delta t_{1,i}$ is the time offset between antenna 1,  to which the other antenna waveforms are referenced,  and antenna $i$.
 
The timing difference between pairs of antennas holds information about the arrival direction of the radio signal and are checked against those calculated from an ice model using uniform index of refraction. 
A pseudo-$\chi^{2}$ is computed for a series of trial source locations in 1-degree bins in $\theta$, $\phi$ and logarithmically spaced bins in radial distance $R$. This is calculated as follows:

\begin{equation}
  \mbox{pseudo-}\chi^{2} = \sum_{i} \frac{(\Delta t_{1,i}^{meas} - \Delta t_{1,i}^{exp})^{2}}{\sigma^{2}}
  \label{eq:analysis:Reconstruction:ChiSq}
\end{equation}
where $\Delta t_{1,i}^{meas}$ is the measured offset between antenna $1$ and antenna $i$, $\Delta t_{1,i}^{exp}$ is the calculated offset expected from a trial source location and $\sigma$ is taken to be 1~ns (which is similar to the timing uncertainty expected given uncertainty on the antenna positions).
The reconstructed location is that which minimizes the pseudo-$\chi^{2}$ and hence corresponds to the most likely physical location given the measured time offsets.

 This method has the benefit of using the rich information contained within the digitized waveforms (correlation techniques result in precision of $\sim 150$~ps resolution in timing differences between pairs of antennas) as well as providing a parameter that describes the goodness of fit in psuedo-$\chi^2$, upon which a cut can be placed. 
A requirement for good reconstruction will also reject a large number of thermal events since they will have essentially random offsets between antennas and the preferred source location will, in general, have a relatively large psuedo-$\chi^2$ value associated with it. The reconstruction of over 2 million events from an in ice calibration source is shown in Fig.~\ref{fig:csw-reconstruction}.

A CSW is formed for the HPol and VPol antennas separately and two parameters are derived that are used to identify neutrino-like signals. 
The first parameter is the peak voltage in the CSW, which acts as a measure of power in the constituent antennas. 
The cross-correlation waveform is computed for each antenna with the CSW of the remaining antennas (since a waveform will be maximally correlated with itself it is not included in the CSW). 
The maximum cross-correlation is found in each of these waveforms and summed to form a variable called `sum of correlation values', which acts as a measure of coherence. 
A linear combination of these parameters is taken to maximize the separation between thermal events and a combination of simulated neutrino and calibration pulser events. 
The resulting cut parameter, dubbed `Powherence', is a measure of both power and coherence requirements between antennas.

Having applied the CW, psuedo-$\chi^2$ and Powherence cuts to the VPol and HPol antennas separately, cuts are made to remove time periods and directions producing large numbers of passing events. The Coherently Summed Waveform analysis places few cuts designed to remove specific backgrounds and instead attempts to remove remaining backgrounds 
through a cut that leverages their repetitive nature, referred to as the ``Good Times" Cut. 
Geometric cuts are made based on reconstruction.

The ``Good Times'' Cut firstly masks off the first and last 30 days of 2011 and 2012 to avoid the peak of human activity at South Pole Station. 
The second stage is to identify, and mask off, days of the year that see a significant number of events passing all but the Geometric Cuts. 
A day is masked off if the total number of passing events in the previous, current and following day exceeds 14 events.

An additional timing cut is included to identify calibration pulser events. 
The Calibration Pulser Timing Cut used in this analysis rejects events occurring within 200~ns of the expected calibration pulser signal arrival time.

\begin{figure}
\centering
\includegraphics[width=3.0in]{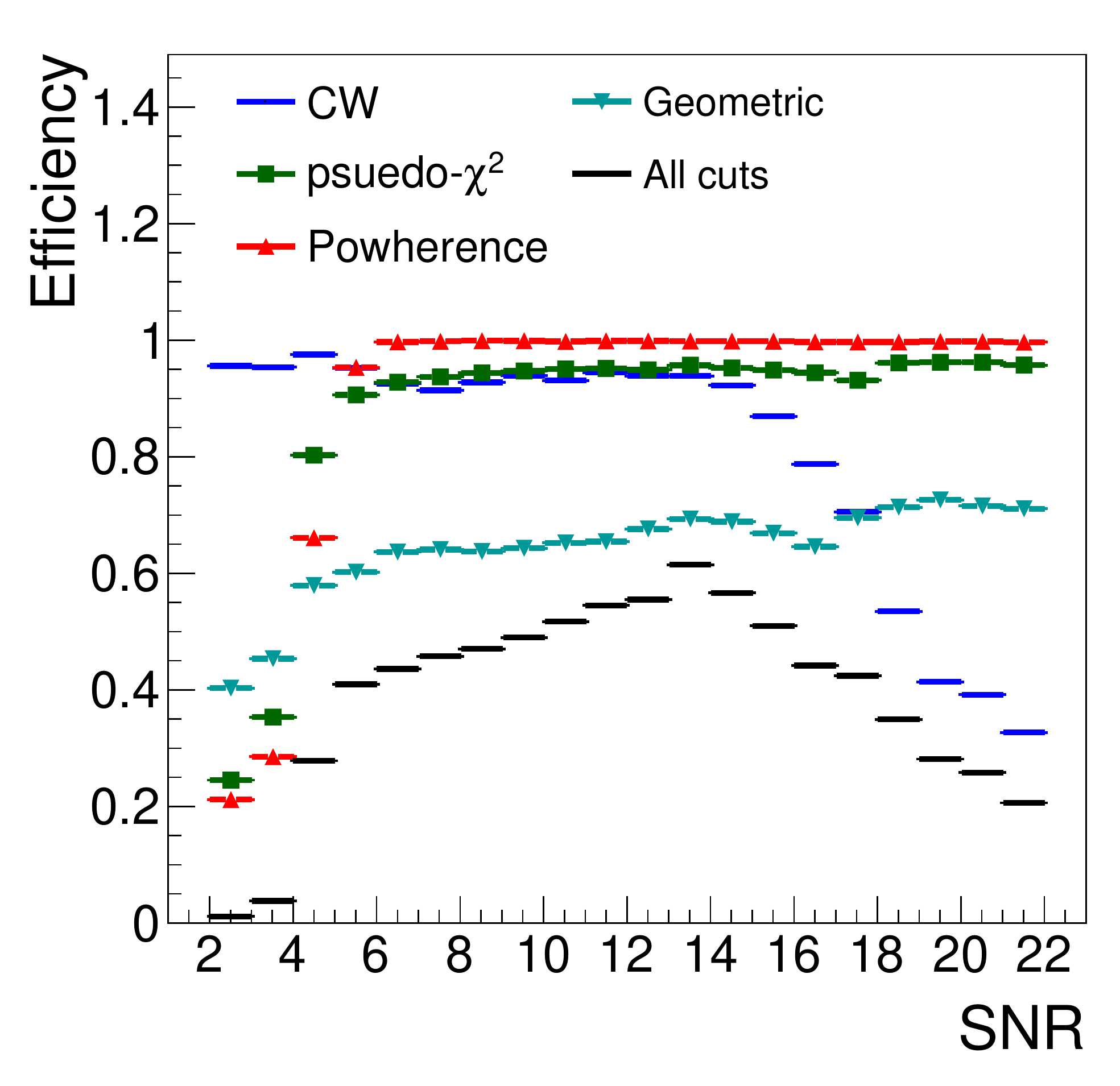}
\caption{\label{fig:cut_efficiencies_UCL} 
The efficiency of each cut compared with the total number
of triggered events as a function of SNR.  Just as for the Interferometric Map Analysis (Fig.~\ref{fig:cut_efficiencies} (a)), this plot does not include the effects of cuts that change the Effective Livetime, i.e. the Good Times Cut and the Calibration Pulser Timing Cut, which rejects 50\% of events independent of signal strength. 
}
\end{figure}

A conservative approach was taken in identifying geometric cuts to remove anthropogenic noise signals. 
The CSW reconstruction achieves $\sim$1~degree resolution for both simulated neutrino events and calibration pulser signals.  
A 50-degree region in azimuth corresponding to the direction of the IceCube Laboratory, as well as 10-degree regions around calibration pulser locations were masked off. 
In addition, events are rejected where the reconstructed source location is above the ice. 
The efficiency for all cuts can be found in Fig.~\ref{fig:cut_efficiencies_UCL}.

Upon application of these cuts to TestBed data taken during 2011 and 2012 results in no events surviving and hence there being no candidate neutrino events.

\subsection {Template-Based Analysis }

The third analysis strategy presented in this paper traces its heritage to the RICE experiment~\cite{PhysRevD.85.062004}, which defined `background' generically as any repetitive waveform or hit antenna pattern. 
In this approach, a sequence of event-selection criteria are initially applied to suppress both anthropogenic and thermal noise relative to `interesting' events (either in-ice neutrino interactions, typically coming from below a given ARA station, or perhaps down-coming radio signals from extensive air showers (EAS)) as follows. 

First, CW contamination is reduced by filtering any CW line which has more than 8\% of the total power 
in the frequency spectrum, and then continuing with the analysis on that filtered event.
Next, triggered events must have at least four antennas with voltage excursions larger than $6~\times$ the root-mean-square voltage 
$\sigma_V$. 
The $\sigma_V$  for a particular antenna is measured using forced triggers (and excluding CW contributions).

\begin{figure*}[!ht]
\centering
\subfloat[]{\includegraphics[width=0.5\textwidth]{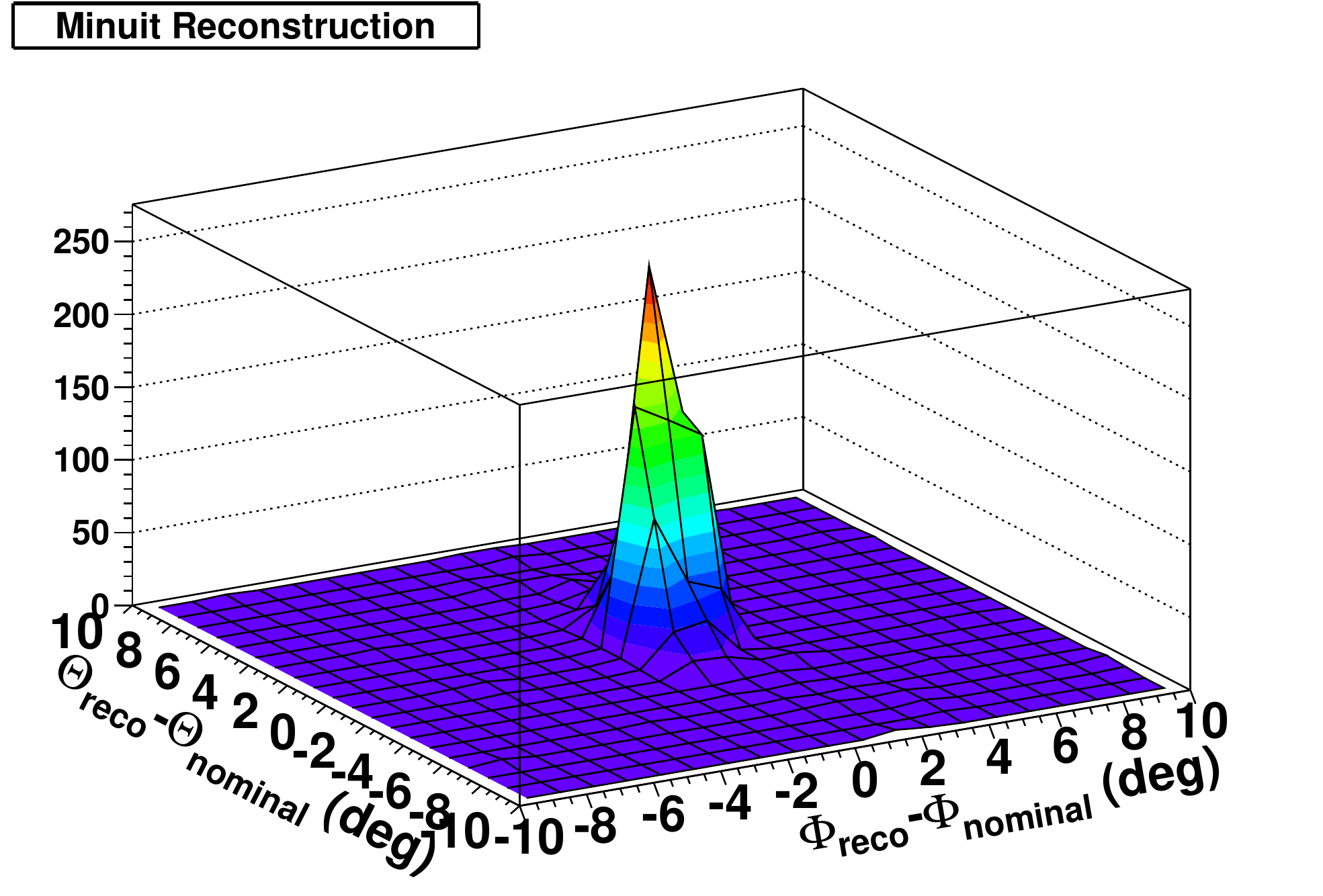}} 
  \subfloat[]{\includegraphics[width=0.5\textwidth]{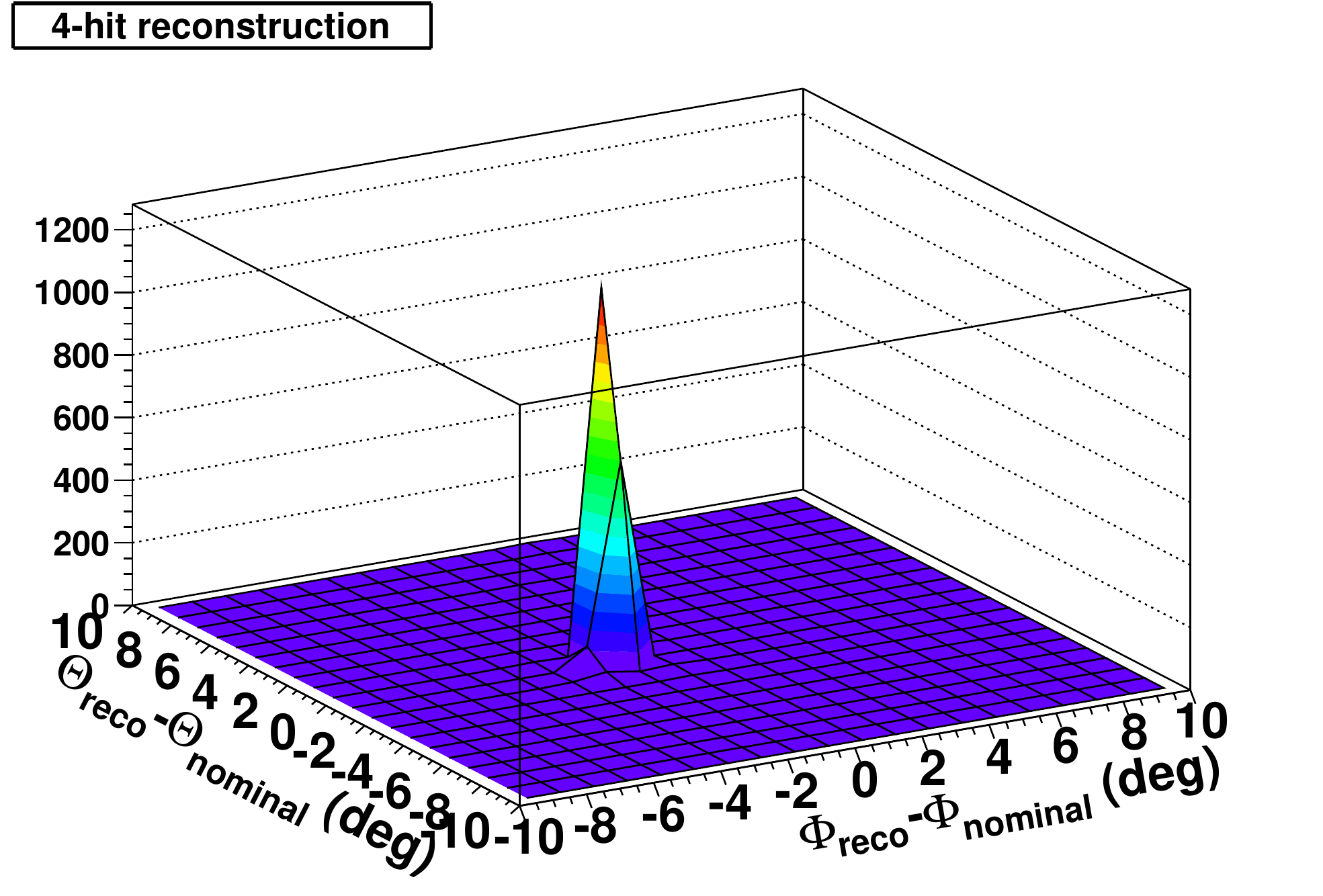}} \\
    \subfloat[]{\includegraphics[width=0.5\textwidth]{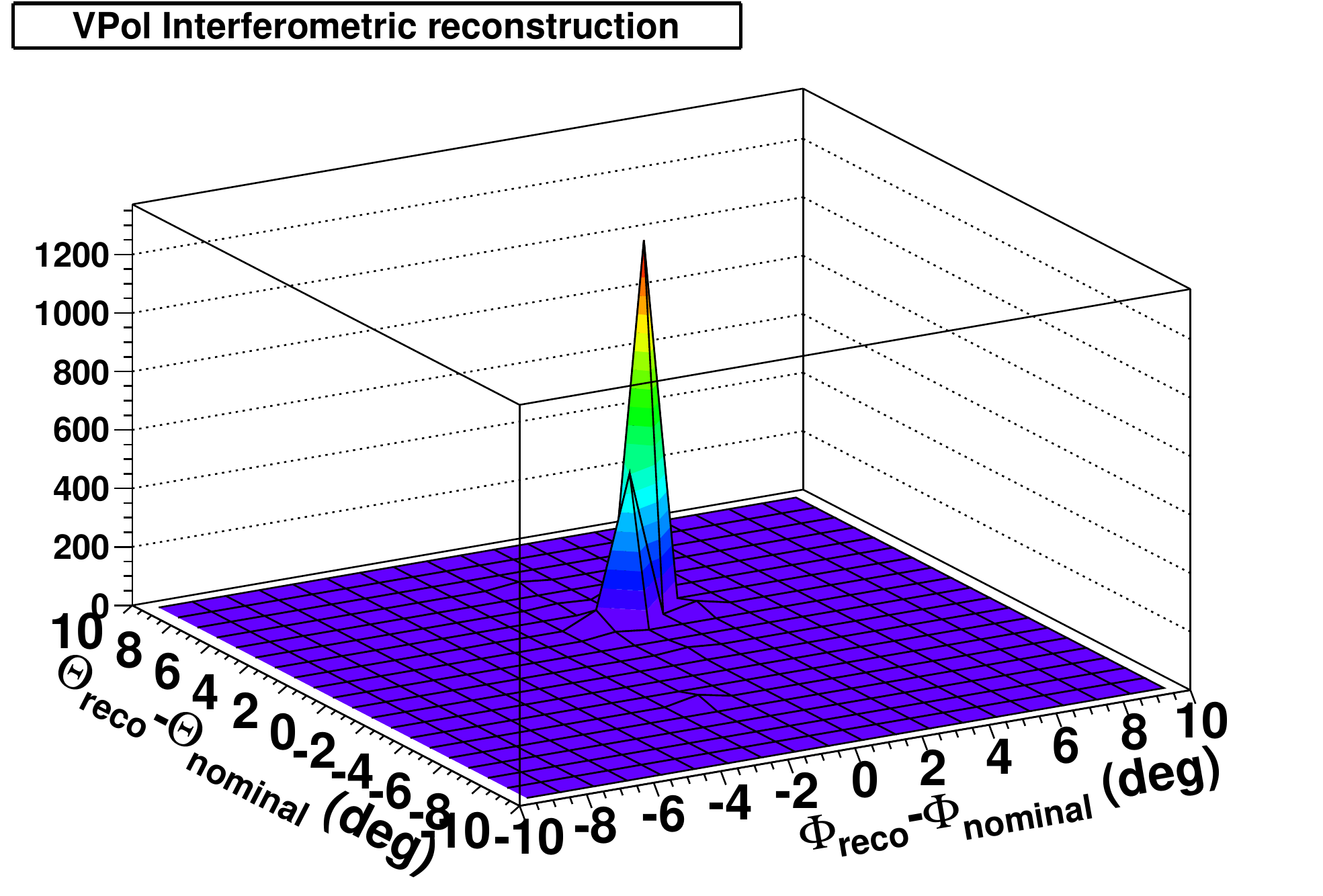}}
      \subfloat[]{\includegraphics[width=0.5\textwidth]{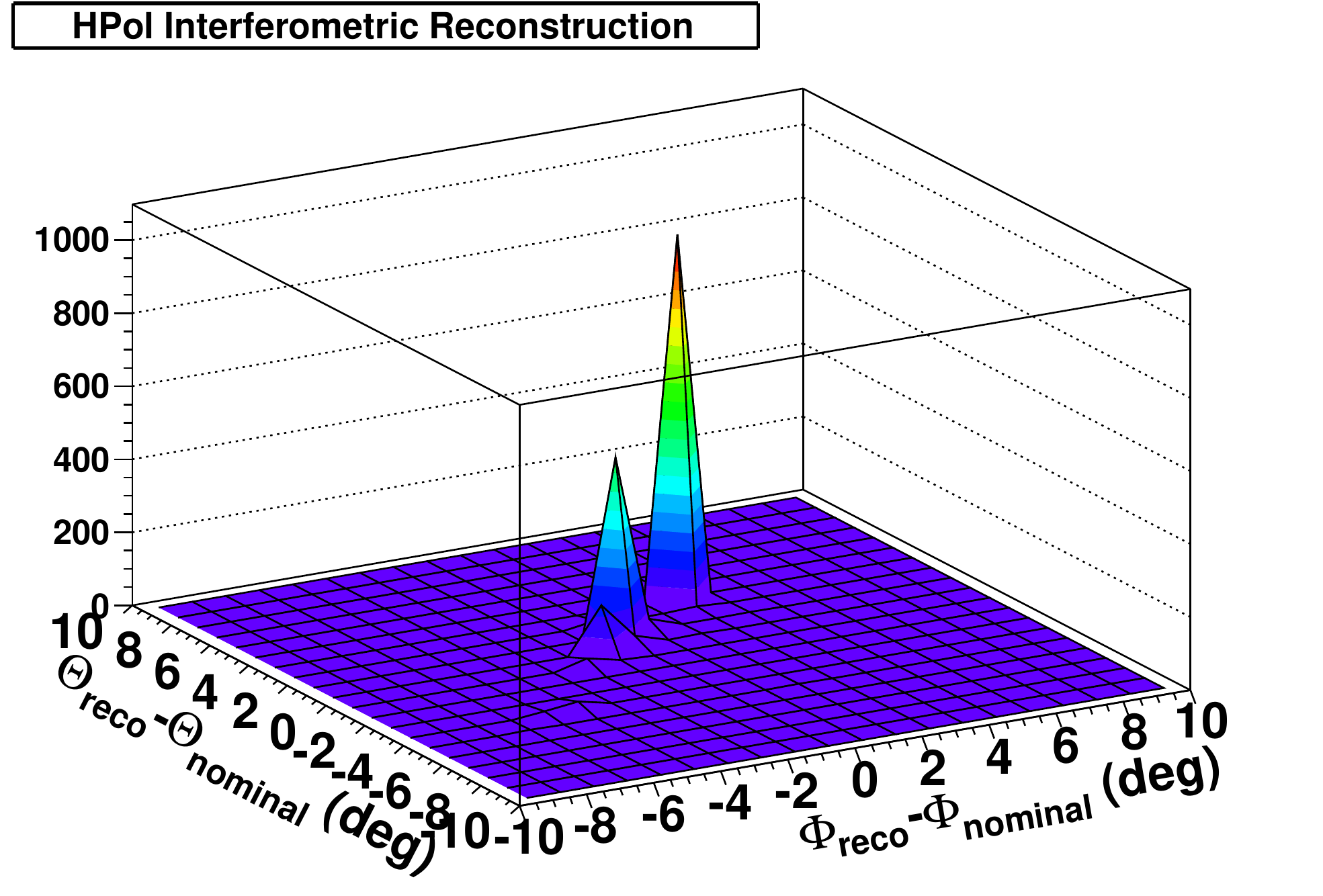}}
\caption{\label{fig:multipanel} Angular reconstruction of the englacial Testbed calibration pulser, from data taken in June, 2011.  Shown is the deviation between the reconstructed polar angle ($\theta_{\rm{reco}}$) and the nominal calibration pulser polar angle ($\theta_{\rm{nominal}}$), in degrees, as well as the corresponding deviation along the azimuth ($\phi_{\rm{reco}}-\phi_{\rm{nominal}}$),  (a) using the ROOT-based Minuit package, (b) solving for the source coordinates analytically for four-hit subsets of all hit antennas (both vertical as well as horizontal polarizations), and taking a weighted average of the found vertex solutions, (c) solving for the source location interferometrically, using only the VPol Testbed receivers, and (d) solving for the source location interferometrically, using only the HPol Testbed receivers.}
\end{figure*}

Second, triggered events must have a well-reconstructed, single source vertex point, as defined by the event $\chi^2$ (defined below), and using source identification algorithms based on RICE codes. 
In this source reconstruction scheme, antennas are assigned a ``hit-time'' corresponding to the time at which the voltage magnitude exceeds 
$6~\sigma_V$. 
The source vertex point ${\bf{r_S}}$ for an event occurring at time $t_S$ is determined in three complementary ways:
\begin{enumerate}
\item Using the CERN-based MINUIT minimization package, 
we find the space point which minimizes the sum of the propagation-time residuals, assuming that vertex point. I.e., we minimize
$\chi^2=\sum\limits_i(t_S-[t_i-|{\bf{r_S}}-{\bf{r_i}}|/c])^2$, where $t_S$ is the calculated propagation time from the putative source point to the $i^{th}$ antenna, $t_i$ is the measured time for that antenna as defined by the first 
$6~\sigma_V$ 
criterion outlined above, $\bf{r_S}$ is the putative source point in coordinate space, $\bf{r_i}$ is the known location for the $i^{\rm{th}}$ antenna, and the sum runs over all the hit antennas. 
\item Second, we find that space point defined as the centroid of the event vertices defined by subsets of four hits of the same polarization. That space point can be thought of as the intersection point of spheres centered on each hit antenna, with a spherical radius 
$r=c(t-t_0)$, 
and $t_0$ the time of the in-ice neutrino interaction. 
\item The results of the previous two calculations are compared against the reconstructed source location using standard ARA interferometric techniques. In fact, the reconstructed angular source resolution for the three techniques are very comparable, as shown in Figure \ref{fig:multipanel}
\end{enumerate}

The third event-selection criterion requires  a total minimum waveform power (defined as $\Sigma(V_i^2)$ for all the in-ice antennas)  to suppress any thermal noise events which pass the four-fold
6~$\sigma_V$ 
requirement. 

Next, if the source location for events passing the previous two requirements is consistent with the known location of the englacial calibration pulser, the event is rejected as a pulser event.

In the final step, triggered events passing the first four requirements are then compared to all other events satisfying those requirements. If the two events are `similar' (as defined by a direct dot product between the two event waveforms, or by the timing pattern of the hit antennas), the events are rejected as `repetitive' and unlikely to arise from `interesting' physics processes such as neutrinos interacting in-ice, or radio waves emanating from charged cosmic ray air showers above the array. 

Application of the above event selection to Testbed data acquired between March 2011 and August 2011 results in one event passing all applied cuts; this event is considered a misidentified in-ice calibration pulser event because
its timing and amplitude characteristics are typical of those events.

\section{Livetime Calculation}
A livetime is obtained for each analysis, defined as the total amount of time covered by the data
set where the trigger is available.
For each analysis, the trigger deadtime is calculated by accumulating the number of 10~MHz clock cycles $N_c$ during one GPS second when the trigger was not available due to the waveform readout process after the trigger or any other issue 
that would cause the Testbed to be unable to trigger.
This counted number of clock cycles $N_c$ gives us the livetime fraction of that second as $1 - N_c/10^7$.
The total livetime of the Testbed is obtained by accumulating the livetime fraction from each second over the entire data set while avoiding double counting of the livetime from same GPS second.
For the Interferometric Map Analysis, if any second has a deadtime $\>10$\%, the entire second is rejected.
The overall livetime 
for the Interferometric Map Analysis 
 from the 2011-2012 Testbed data is 56.8\%, or 415~days.

Since some of analysis cuts are defined to reject certain periods of activity, one can also define an Effective Livetime for comparison between the analysis methods.
For the Interferometric Map Analysis, there are three such cuts that reduce the Effective Livetime. 
These cuts are the Calibration Pulser Timing Cut, the IceCube Drilling Season Cut, and the Good Baseline Cut, all described earlier.
After all of these cuts are considered, the Effective Livetime is 224 days.
For the Coherently Summed Waveform analysis, two cuts are applied to find the Effective Livetime: a Calibration Pulser Timing Cut and the ``Good Times" Cut, both described above.
After considering this cut, the Effective Livetime for the Coherently Summed Waveform Analysis is 206 days.

\section{Results}

No neutrino candidate events were found for the Interferometric Map Analysis and the results from this analysis are used to derive constraints on the neutrino flux.  
Compared to the Interferometric Map Analysis, the Coherently Summed Waveform Analysis has a 30\% higher analysis efficiency and a $\sim10\%$ lower effective livetime, thus limits derived from the latter give a very similar result.
We find it quite encouraging that these two complementary analyses give such similar results.  
We note that they both use AraSim to interpret the results as limits.

The effect of the successive cuts in Stage 2 of the Interferometric Map Analysis is summarized in Table \ref{tab:events}. 
After the Event Quality and the Reconstruction Quality Cut are applied, for this table the events are examined in HPol and VPol channels separately.
While a single event can pass the HPol and VPol Reconstruction Quality Cut simultaneously and be considered in both channels, only a small number of events ($\sim 100$) did so.

\begin{table*}[t]
    \begin{tabular*}{\textwidth}{@{\extracolsep{\fill} } l  | c  | c | c  | c | c | c} \hline \hline
Total & \multicolumn{6}{c}{ 3.3E8} \\ \hline
Cut & \multicolumn{6}{c}{Number passing (either polarization)} \\ \hline 
Event Qual. &  \multicolumn{6}{c}{1.6E8}  \\
Recon. Qual. &  \multicolumn{6}{c}{3.3E6} \\ \hline
  & \multicolumn{3}{c}{ VPol} & \multicolumn{3}{c}{ HPol} \\ \hline
  & \multicolumn{3}{c}{Rejected} &  \multicolumn{3}{c}{Rejected} \\
   & In sequence & as last cut & as first cut & In sequence & as last cut & as first cut \\ \hline
  Recon. Qual. & 1.8E6 &  & & 1.4E6 & &  \\ \hline
  SP Active Period & 1.4E6 & 125 & 4.9E5 & 1.1E6 & 13 & 3.5E5\\ \hline
  Deadtime $<0.9$ & 1.4E6 & 0 & 3.2E4 & 1.1E6 & 0 & 9.2E3 \\ \hline
  Saturation & 1.4E6 & 0 & 1.4E4 & 1.1E6 & 0 & 618 \\ \hline
  Geometric, except SP & 1.3E6 & 7 & 9.9E4 & 1.0E6  & 0 & 4.6E4 \\ \hline
  SP Geometric & 1.1E6 & 0  & 2.9E5 & 9.0E5  & 1  & 2.0E5 \\ \hline
  Gradient & 1.1E6 & 0 & 1.4E4 & 9.0E5 & 0 & 4.6E3 \\ \hline
  Delay Difference & 1.8E5 & 0 & 1.5E6 & 1.5E5 &0 & 1.2E6\\ \hline
  CW & 1.8E5 & 0 & 1.3E4 &  1.4E5 & 1 & 3.4E4\\ \hline
  In-Ice & 1.7E4 & 15 & 1.6E6 & 1.9E4 & 1 & 1.2E6 \\ \hline
  Peak/Corr & {\bf 0} & 1.7E4 & 1.8E6 & {\bf 0} & 1.9E4 & 1.4E6\\ \hline \hline
\end{tabular*}
\caption{\label{tab:events} This table summarizes the number of events passing each cut in the Interferometric Map Analysis,
 in Phase 2 (2011-2012, excluding Feb.-June 2012).
We list how many events each cut rejects as a last cut,
and how many are rejected by each cut if it is the first cut.  After the
Event Quality and Reconstruction Quality Cuts are applied, VPol and HPol
and considered as two separate channels for the purpose of tabulation, independent of one another.}
\end{table*}

After finding no neutrino candidate events passing all cuts, we set limits on the neutrino flux given the effective volume of the Testbed derived from AraSim and total livetime of the period examined.
The effective volume,  $V_{\rm{eff}}$, is found for each energy bin by simulating a large number ($\sim10^6$) of events~\cite{Kravchenko:2006qc}:
\begin{equation}
V_{\rm{eff}} = 
\frac{V_{\rm{cylinder}}}{N}\sum\limits_{i = 1}^{N_{\rm{passed}}}{w_{\rm{i}}}.
\end{equation}
Each event is given a weight $w_{\rm{i}}$ equal to the probability that the neutrino was not absorbed in
the earth, given its direction and position of the interaction.
Then $\sum\limits_{i = 1}^{N_{\rm{passed}}}{w_{\rm{i}}}$ is the weighted sum of the number of events 
that triggered and passed all analysis cuts, and $N$ is the total number of events thrown.
The neutrino interactions are thrown in a cylindrical volume centered around the detector, denoted
$V_{\rm{cylinder}}$.

To find the 90\% confidence level limits on the differential flux, we estimate for each decade in energy the upper limit on the number of events predicted in that energy decade given that no events
were observed.  For each energy bin,
\begin{equation}
2.3 =  \frac{dN(E)}{d\log_{10}(E) dA dt d\Omega}  d\log_{10}(E) ~dA(E) ~dt ~d\Omega
 \end{equation}
 where $dN(E)/d\log_{10}(E)/dA/dt/d\Omega$ is the number of events per area per time per steradian arriving at earth in each decade in energy.
 The factor of 2.3 is the 90\%  confidence level upper limit on the number of events expected assuming a Poisson distribution and with zero events observed.  Note that one decade in
 energy is always the bin size chosen
 for this calculation regardless of the number of points plotted for the differential limit.
 
We take $dt=T$, where $T$ is the livetime of the examined period, and $d\Omega=4\pi$. 
 For the area, we use the thin target approximation, meaning that the dimensions of the detector are much smaller than the
interaction lengths~\cite{Williams:2004bp}:
\begin{equation}
A_{\rm{eff}} (E) \approx
\frac{V_{\rm{eff}}(E)}
{l_{\rm{int}}(E)}
\end{equation}
where $l_{\rm{int}}$ is the interaction length.
 Then if we substitute $dN(E)/d\log_{10}(E)= E ~dN(E)/dE ~\ln(10)$
 using $d\log_{10}(E)=d\ln(E) / \ln(10)$, and take the differential flux to be $F(E)=dN(E)/dE/dA/dt/d\Omega$,
 we find:
\begin{equation}
E~F(E) =
\frac{1}
{4\pi T A_{\rm{eff}}(E)}
\frac{2.3}
{\rm{ln}(10)}
\end{equation}

\begin{figure}
\includegraphics[width=3.0in]{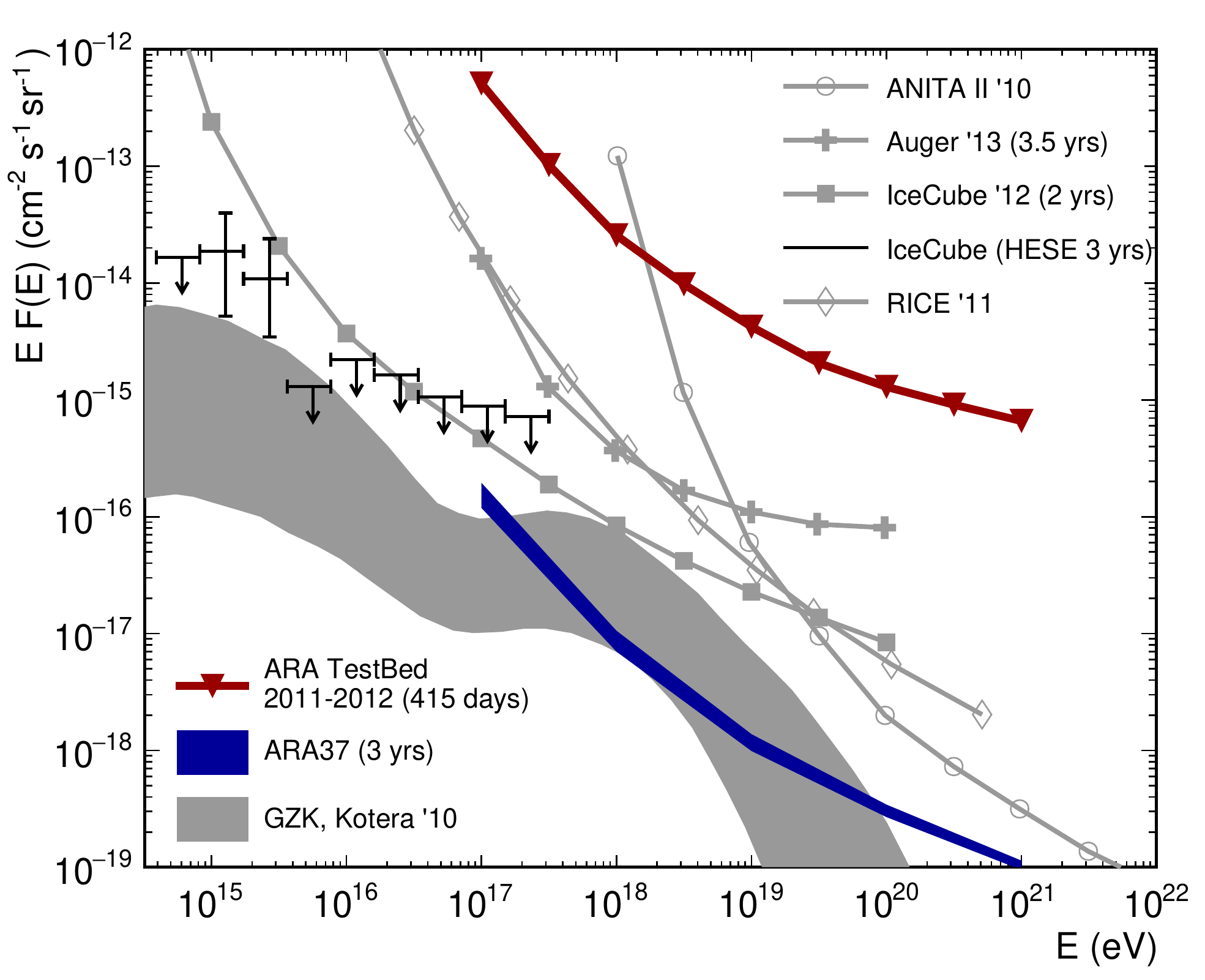} 
\caption{The limits set by this analysis compared with the projected ARA37 trigger-level sensitivity and results from other experiments ~\cite{Gorham:2010xy,Abreu:2013zbq,Aartsen:2013dsm,Aartsen:2014gkd,Kravchenko:2011im}. 
These Testbed limits are from the Interferometric Map Analysis, although the Coherently Summed Waveform Analysis gives a similar result.}
\label{fig:limit}
\end{figure}

The limit curve shown in Fig.~\ref{fig:limit} was made for the Interferometric Map Analysis
although the Coherently Summed Waveform Analysis produces very similar results, as described earlier.
Fig.~\ref{fig:limit} can then calculated from:
The projected limits for ARA37 shown in the same figure are derived from trigger-level sensitivities only, with 100\% analysis efficiencies assumed for simplicity.

To understand the impact of the choice of flux model for the optimization of the cuts, we also optimized the Peak/Correlation Cut using a flux model from Ahlers 2010~\cite{Ahlers:2010fw} (best fit model using extragalactic cosmic ray crossover energy $\rm{E_{min}}=10^{18.5}$~eV case).
With this flux model in place of the Kotera maximal model, we again found an optimal Peak/Correlation Cut Value of 8.8, the same result from Kotera maximal model.

\section{Sensitivity to Cosmic Ray Events} \label{sec:EAS}

Radio emissions from extensive air showers have been predominantly studied in the frequency range $f<$100 MHz, corresponding to the regime for which the signal strength is largest. Nevertheless, as demonstrated by the ANITA experiment, air showers also radiate sufficient signal in the $200-1000$~MHz band to be easily observed above thermal background for air shower primaries exceeding 1 EeV at $\sim$100 km distances from shower core to receiver. The ARA receiver array acceptance differs from ANITA in three main respects: 1) the geometric acceptance is considerably smaller, corresponding to only those cases for which the incident air shower impacts within an approximately one-degree wide annulus with an appropriate incidence angle such that the signal refracts down to the ARA receiver array, 2) whereas the ANITA horn antennas have excellent response to the predominantly HPol geomagnetic signal, the ARA antennas are viewing the incident signal close to the null of the antenna beam pattern, 3) the ARA antennas are considerably closer to the shower core, implying (in principle) a lower cosmic ray energy threshold.

\begin{figure}
\includegraphics[width=0.5\textwidth]{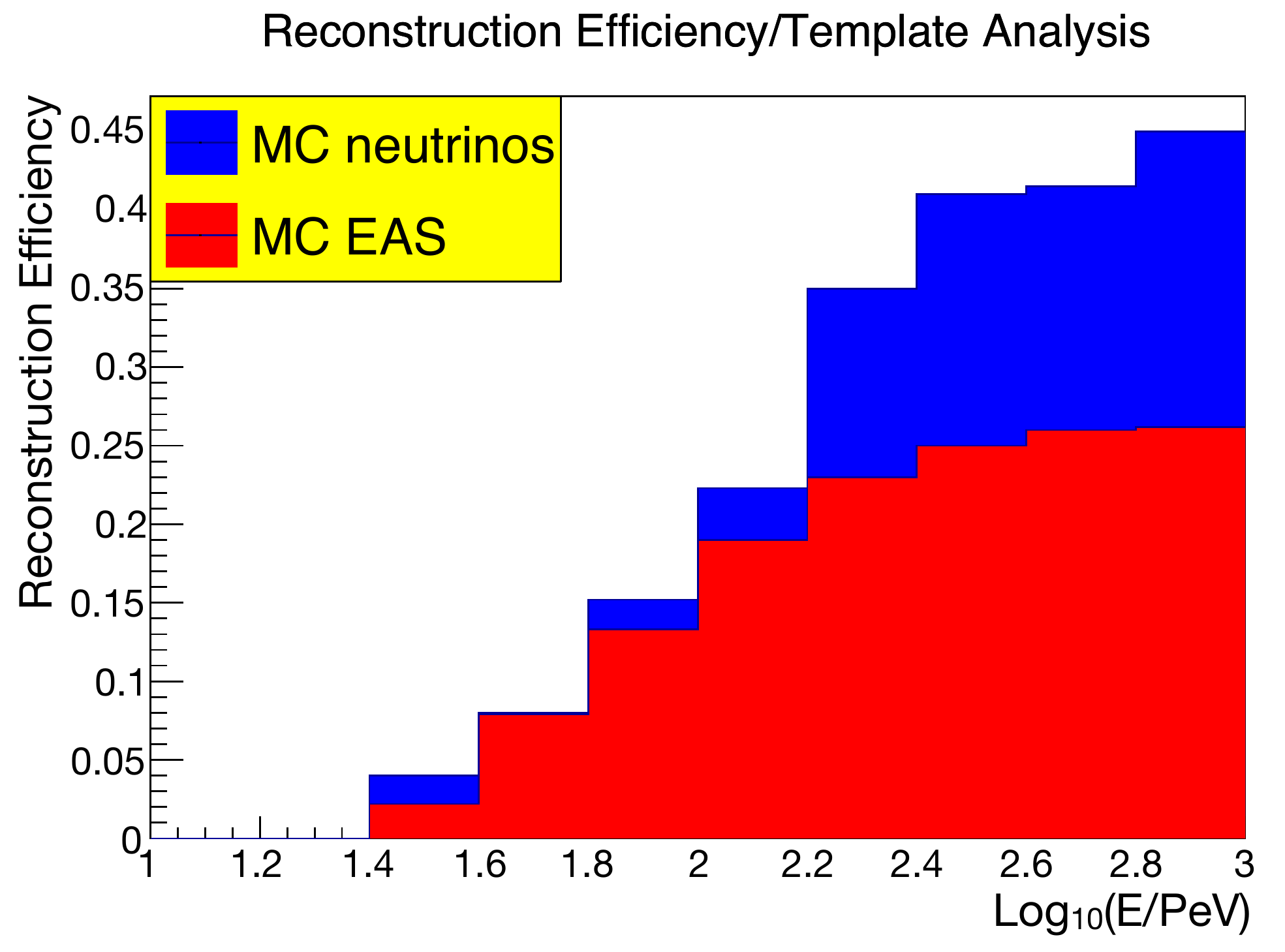}
\caption{The reconstruction efficiency of the Template-Based Analysis as a function of the energy of primary particle from RA-RA neutrino simulation and EAS cosmic ray air shower simulation sets.
}
\label{fig:RARA_EAS}
\end{figure}

We have estimated the expected sensitivity of the ARA testbed to extensive air showers using a sample of 3000 simulated events, using the CoREAS (Corsika + REAS) Monte Carlo package, as follows:
\begin{enumerate}
\item The frequency domain signal, over the RF regime ($f<$1000 MHz) is calculated for each air shower.
\item The signal strength, for all frequencies, is then reduced by the appropriate transmission coefficient at the air-ice interface, neglecting any surface roughness effects.
\item The RF signal penetrating into the ice is now further attenuated by ice RF absorption (this is a minor effect).
\item The signal is now `dotted' into the antenna response (gain and polarization), as a function of frequency.
\item The resulting signal is now inverse Fourier transformed into the time domain and superimposed upon `forced trigger' ARA testbed events, taken to be representative of the ambient noise environment. If the magnitude of the resulting signal voltage exceeds $4~\sigma_V$, the simulated antenna is considered to be ``triggered'' at the lowest trigger level. Three triggered antennas constitute a triggered event.
\item Finally, the simulated triggered events are analyzed as real data.
\end{enumerate}
This procedure results in the RA-RA efficiency vs. EAS primary energy $\epsilon$(E) curve in Fig.~\ref{fig:RARA_EAS}. To estimate the total number of expected events, the $\epsilon$(E) curve shown is folded in with the well-known primary cosmic ray charged spectrum. Accounting for the livetime, we expect less than 0.5 EAS events over the term 2011-2012, consistent with observation.

For each pair of antennas, a time delay between signal arrival times corresponds to a ``ring" on the interferometric map, and among a set of multiple antennas, the rings from all pairs should intersect at the ``true" direction of origin of the signal.
Sometimes, rings can overlap at other points on the map and this can result in a misreconstruction.

We estimate the fraction of cosmic rays that might be misreconstructed and identified as coming from the ice in both data and simulation.
Here we impose cuts from the Interferometric Map Analysis, which is more restrictive than the Template Analysis.
Both analyses favor high signal-to-noise events.

To investigate the percentage of time that a down-going cosmic ray can be misreconstructed and point to the ice, we removed our ``SPS" Geometric Cut, the Geometric Cuts on repeating locations and the In-Ice Cut but kept all other analysis cuts the same.
This leaves 42 events (from the 10\% set), of which none were misreconstructed as coming from the ice.
If we loosen the Peak/Correlation Cut Value we found that 1.3\% and 8.3\% of events were misreconstructed as coming from the ice when the cut value is loosened from 8.8 to 7.0 and 6.5 respectively.

To verify these rates of misreconstructions to the ice with AraSim, simulated calibration pulser signals are propagated from positions above the ice and reconstructed using the Interferometric Map technique.
When we impose the same set of cuts as we did in the data for this study, we do not find any simulated events that reconstruct to the ice out of a sample of $\sim10^5$ covering all incident angles above the horizon.
When the Peak/Correlation Cut Value is loosened, the rate of misreconstructions to the ice reproduces the same qualitative behavior as is seen in the data, but we find about half as many misreconstructions to the ice in the simulated set as we find in the Testbed data set.
We conclude that for a down-going impulsive event, the probability of a miscreconstruction to the ice is conservatively less than 1\%.

Based on a maximum of 0.5 estimated air shower events and a $<$1\% chance of misreconstruction to the ice, we expect $<$0.005 cosmic-ray events, compared to the $0.03 ^{+0.26} _{-0.03}$ total background events.  
Since cosmic ray air showers have never been observed by an in-ice array, we stress that we have purposely built in conservative assumptions into our estimates in order to allow for unevaluated systematic errors.

\section { Systematic Uncertainties}

In this section, we will discuss the systematic uncertainties in the Interferometric Map Analysis.
We considered systematic uncertainties in both the background estimation and analysis efficiency.
While the uncertainty on our background estimation is derived solely from the errors on the best fit line used to
extrapolate the background estimate described in Section~\ref{sec:IMA}, for the analysis efficiency we consider the effect of 
the antenna model, ice index of refraction model, ice attenuation model, and neutrino cross section model.
Each systematic error is obtained by changing only one parameter value at a time from our basis and estimating the impact on the result from each.

The systematic uncertainty on the background estimation is derived from the 
errors on the best fit  exponential  function used in the extrapolation described in Section~\ref{sec:IMA}.
Recall that the best fit to $N_{\rm{diff}}=e^{a\cdot x+b}$ gave
$a = -4.29\pm0.26$ and $b = 31.70\pm1.67$.
We moved each fit parameter alone by one standard deviation in both the positive and negative directions, and obtained the maximum deviation in the background estimate in each direction.
We find the number of background events to be
0.03~$^{+0.26}_{ -0.03}$~events in the 90$\%$ data set in the Stage~2 analysis.

Modeling the expected frequency-dependent phases of the neutrino-induced measured pulses 
contributes an important  systematic error to our analysis efficiencies.
We use two different techniques for modeling the phases, and they 
put upper and lower bounds on our limit and projected sensitivity for ARA37, which are
based on our current trigger and
given the effects so far included in our simulations.  Conservatively, our main result in this paper (the
red line in Fig.~\ref{fig:limit}) uses
our default model, while we believe that a more accurate model would give an improved limit.

We model the received impulse (including frequency-dependent phases) from each neutrino interaction using two different methods, the first being the default and the second for comparison.
The most important difference between the two models is in their frequency-dependent phases that come about from the RF emission itself and convolved with the response of the antennas and electronics.
We believe that the ``true'' impulse would give a result that is in between the two.
The first, default approach models the frequency-dependent phase of the RF emission, as well as the phase response of antennas, filters and amplifiers as described in Section~\ref{sec:simulations}.
The second model for the phase, used for comparison, is quite simplistic, with the phase of the RF emission being +90$^\circ$ for positive frequencies and -90$^\circ$ for negative frequencies, and the phase response of antennas and electronics being flat (having no impact).

While the second, simple model of the phase response produces a received pulse that is too narrow,
we have found by comparing simulated and measured calibration pulser waveforms that
our default method simulates pulses that are too broad.  
The narrow pulses from the second method result in analysis efficiencies
that are too high, and the broader pulses from the default method fail our cuts more often than they should.
This excessive broadening of the pulse is believed to be dominated by the antenna response model, and
future measurements of the phase response of our antennas are expected to 
greatly reduce this systematic uncertainty.

The second model gives a trigger level sensitivity that is approximately $65\%$ larger than the first model at $10^{17}$~eV, $50\%$ larger at $10^{18}$~eV and $20\%$ larger at the highest energy simulated, $10^{21}$~eV.
At low energies, the dispersion of the signal has a more dramatic effect on the trigger efficiency whereas at higher energies the dispersion has less of an effect due to the the strength of the signal.

The choice of model for the depth dependence of the index of refraction in the firn, both for event generation
and for event reconstruction, provides another source of 
systematic uncertainty in our analysis efficiency since it determines the path taken through the ice and the
arrival direction at the antennas, and also impacts the interferometric maps that are used in analysis.
By default, we used the exponential fit function for the 
index of refraction as a function of depth:
\begin{equation}
n(z) = 1.78 - 0.43 \cdot e^{-0.0132\cdot z}
\label{eq:exp_original}
\end{equation}
where $n$ is the index of refraction and $z$ the depth of ice (positive value for deeper location).
We used two alternative index of refraction models, one using the same format with slightly different parameter values and another with a different functional form. The former of these two alternatives was an exponential fit function:
\begin{equation}
n(z) = 1.79 - 0.43 \cdot e^{-0.013446\cdot z}.
\label{eq:exp_new}
\end{equation}
The latter was an inverse exponential fit function:
\begin{equation}
n(z) = 1.0 + \frac{0.78}{ 1+ e^{-0.023\cdot z}}.
\label{eq:inv_exp}
\end{equation}

All three functions give an index of refraction of $\sim 1.35$ at the surface and $\sim 1.78$ in deep ice and give satisfactory empirical fits to the RICE measurements ~\cite{2004JGlac..50..522K}.
The exponential models have a more shallow firn layer ($\sim 200$~m) compared to the inverse exponential model ($\sim 250$~m), and thus show a more dramatic change in index of refraction as a function of depth.

As the Interferometric Map Analysis bases its reconstruction method on a particular model of index of refraction, this technique thus assumes perfect knowledge of the index of refraction.
In order to estimate the systematic error due to imperfect knowledge of the depth-dependence of the index of refraction of the ice, we try using each of the three models for event generation and/or reconstruction, giving nine combinations including the default combination, and quantify the systematic error as the largest excursions from the baseline result in either direction.
We find the efficiency can only decrease by up to 11.3\% 
at $E_{\nu} = 10^{18}$~eV compared to the default 
due to imperfect knowledge of the depth-dependence of the index of refraction in ice. 
No increase in efficiency was observe for any combination of index of refraction models for event generation and reconstruction.

Likewise, we assess the systematic uncertainty due to in-ice field attenuation by comparing our
our result when two different models are used.
The default model uses a South Pole temperature profile from~\cite{Price} folded in with a relationship between field attenuation length and ice temperature given in~\cite{Matsuoda} as
used in ANITA simulations and described in~\cite{fenfang}.
The alternative ice attenuation length model is based on the ARA Testbed measurement from IceCube deep pulser events published in~\cite{Allison:2011wk}.

By default, our modeling of the effect of ice attenuation is based on ANITA simulations~\cite{fenfang},
where profiles of ice attenuation vs. depth are considered, and for each event, the ice attenuation length 
 is averaged over depth
from the neutrino-ice interaction location to the surface and the result is 
denoted $\langle L_{\rm{atten}}(z) \rangle$, where $z$ is the depth of neutrino-ice interaction location.
The attenuation length is assumed to remain a constant $\langle L_{\rm{atten}}(z) \rangle$ over 
the entire path of the ray in the ice, and
an ice attenuation factor,
\begin{equation}
F_{\rm{IceAtten,default}}^{\rm{total}} = e^{D_{\rm{travel}} / \langle L_{\rm{atten}}(z) \rangle },
\end{equation}
is then applied to the electric field.
Here, $F_{\rm{IceAtten,default}}^{\rm{total}}$ is the ice attenuation factor and $D_{\rm{travel}}$ is the ray travel distance between the neutrino-ice interaction location and the antenna.

The second calculates the total attenuation factor every few 10~m along the path of the ray for each event,
and uses the ice attenuation lengths measured by the ARA Testbed~\cite{Allison:2011wk}.
The total ice attenuation factor from this method is then:
\begin{equation}
F_{\rm{IceAtten,alter}}^{\rm{total}} = \prod _{i=1}^{N} e^{D_{\rm{i}} / L_{\rm{atten}}(z_{i}) }
\end{equation}
where $F_{\rm{IceAtten,alter}}^{\rm{total}}$ is the ice attenuation factor to the electric field strength from the alternative model, $N$ the total number of ray tracing step from neutrino-ice interaction location to the antenna, $D_{\rm{i}}$ the ray travel distance for the corresponding ray trace step $i$, and $L_{\rm{atten}} (z_{i})$ the ice attenuation length at corresponding ray tracing step's depth $z_{i}$.
Due to the fact that the second technique gives us longer attenuation lengths near the surface, it gives a $\sim 10\%$ larger efficiency at $E_{\nu} = 10^{18}$~eV compared to the default model.

Finally, we estimate the uncertainty due to our $\nu-N$ cross section model.
The $\nu N$ cross section model in our simulation is from Connolly $\textit{et al.}$~\cite{Connolly:2011vc} which gives us the central values and upper and lower bounds for the $\nu-N$ cross section as a function of $\nu$ energy.
At $E_{\nu} = 10^{18}$~eV, the uncertainty in the $\nu-N$ cross section give us up to $\sim 30\%$ variance from the central value  (from the upper bound on the neutral current cross section).
There are two competing effects in play when the cross section is increased.  The Earth screening effect leads to a decrease in the number of events reaching the detection volume from below,
while there in an increase in the number of neutrinos that interact once they reach the ice. 
At $10^{18}$~eV, using a $\nu N$ cross section at the lower bound gives a $\sim 6.2\%$ higher neutrino efficiency due to reduced absorption in the earth, while using the upper bound on the $\nu N$ cross section gives a negligible change in efficiency in comparison to the baseline model.

\begin{table}
\begin{center}
\begin{tabular*} {0.35\textwidth}{ | @{\extracolsep{\fill} }  l | c | c |} \hline
Systematic  & & \\ 
uncertainties at $10^{18}$~eV & $+$ (\%) & $-$ (\%)   \\ \hline
Index of Refraction &  0 & 11.3  \\ 
Ice Attenuation Length &  10.2 & - \\ 
$\nu$N Cross Section & 6.2 & 0  \\ 
Phase Response & 50.9 & - \\
Total & 52.3 & 11.3   \\ \hline
\end{tabular*}
\caption{\label{tab:systematics} Summary of systematic uncertainties on the neutrino efficiency
at $10^{18}$~eV. To find these values, we determined the effective volume using different models for the respective parameters. We then found the maximum deviation of these values in either direction from the effective volume using the default parameters.
}
\end{center}
\end{table}

Overall, we estimate that we expect
0.03 $+$ 0.26 $-$ 0.03 
background events with $+52.3\%$ and $-11.3\%$ uncertainties on 
our neutrino efficiency.

\section{Projections for ARA3 and ARA37}
\begin{figure*}[!ht]
\centering
\subfloat[]{\includegraphics[width=0.45\textwidth]{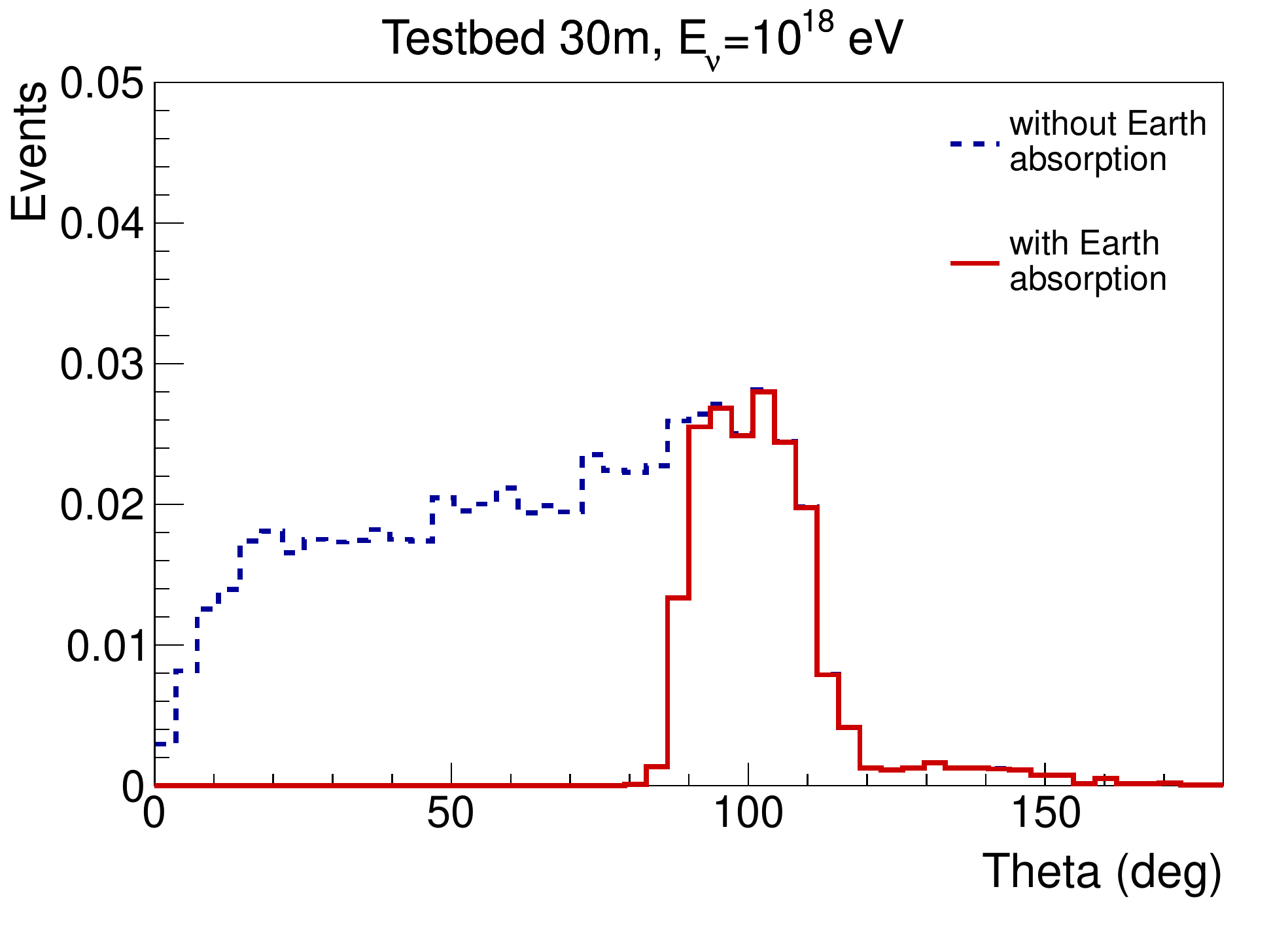}} \hfill
  \subfloat[]{\includegraphics[width=0.45\textwidth]{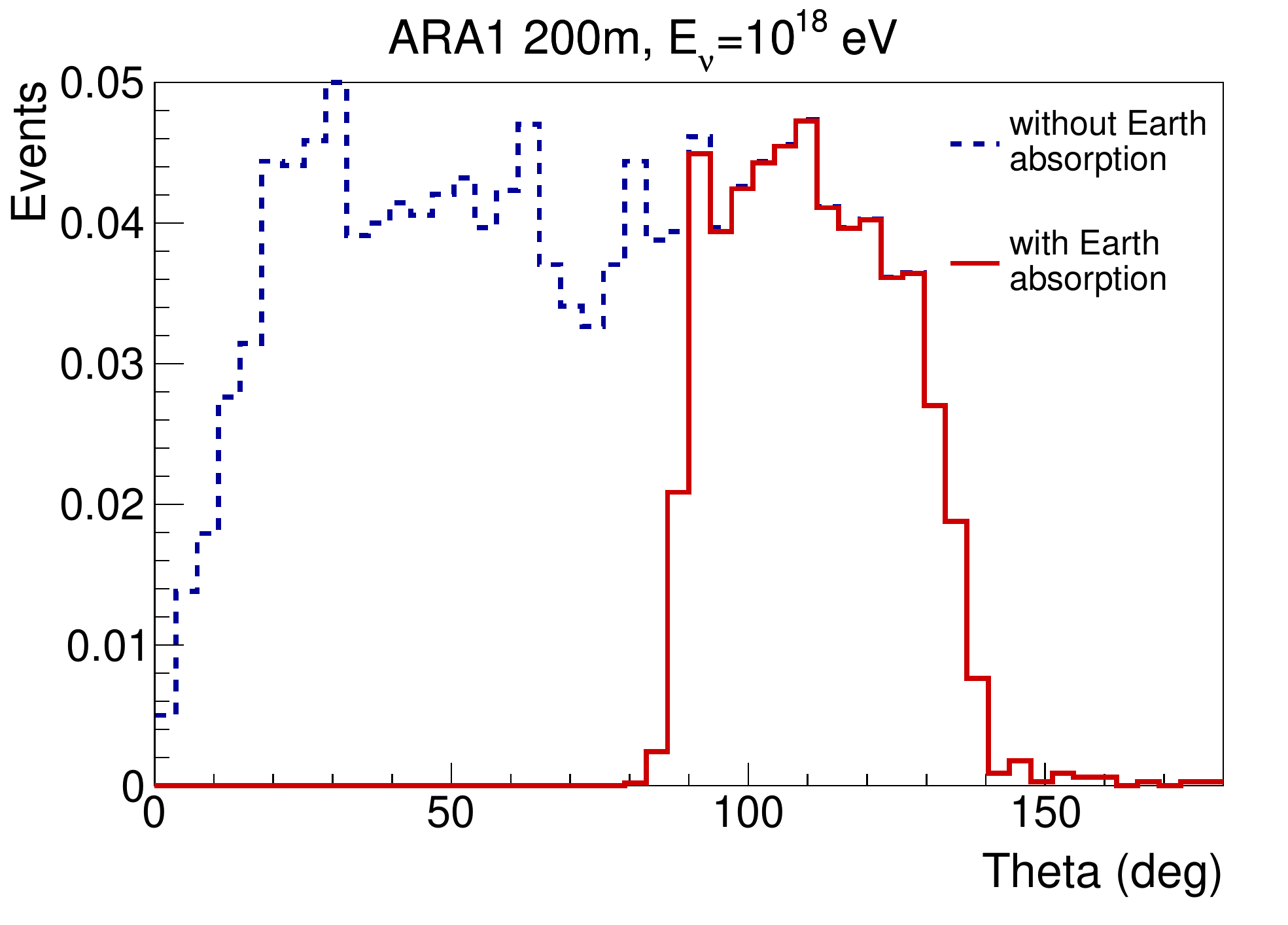}}
\caption{\label{fig:AngularShadowing} These figures show the distribution of zenith angles of incident momenta for simulated neutrinos at $10^{18}$  eV that pass the trigger in AraSim for (a) Testbed at 30~m and (b) a design station at a depth of 200~m. Events on the left side of the plot come from up-going neutrinos with respect to
the south pole, while events on the right come from down-going neutrinos.  The viewable arrival direction zenith angles are generally limited to less than $120^{\circ}$ for the Testbed and less than $150^{\circ}$ for a design station. This limited range of observable arrival directions is due to the combination of the limited viewing region seen in Fig.~\ref{fig:RayTrace} and the requirement that the coherent signal is emitted near the Cherenkov angle, which is relative to the arrival direction. When one adds in the screening effect of the Earth (red lines), almost all events with zenith angles less than $90^{\circ}$ disappear as well and thus the observable range of zenith angles is limited by the geometry of the Testbed by a factor of about 2.}
\end{figure*}

The ARA collaboration aims to build an array of 37 stations to gain enough sensitivity to measure of order 100 UHE neutrinos and exploit the physics and astrophysics information that they carry.
In this section we illustrate the factors that bring us from the sensitivity of this Testbed analysis 
to the expected ARA37 trigger sensitivity in Fig.~\ref{fig:limit}.
For future detector configurations we compare sensitivities at the trigger level only.
We do not know what our analysis efficiencies will be at those stages but expect that they will improve (see below).

Table~\ref{tab:toara37} lists the factors that bring the Testbed sensitivity in this paper to that expected for an ARA37 array at $E_{\nu}=10^{18}$~eV, where, for many cosmogenic neutrino flux models, we expect to measure the largest number of neutrinos.  The factors in Table~\ref{tab:toara37} are all derived from the AraSim simulation and we use the effective area$\times$ solid angle 
$A\Omega_{\rm{eff}}$ as the figure of merit to compare the sensitivity of the detector at different stages.  
The effective area at the Testbed trigger level is a factor of 10 higher than at the analysis level at $10^{18}$~eV.  
Going from the Testbed trigger level to an ARA deep station, we find a factor of 2.8 improvement in sensitivity.  
This is both because a shallow station is limited in the angle of incident RF emission that it can observe,
and also because neutrinos steep enough to produce observable RF emission are subject to more earth absorption (see Fig.~\ref{fig:AngularShadowing}).
From one ARA station to ARA37, the sensitivity scales as the number of stations since at these energies, events tend to be seen by only one station and each station serves as its own independent detector.

\begin{table}
{\bf  From the Testbed to ARA37 at $10^{18}$~eV}
   \begin{tabular*}{3.0in}{ @{\extracolsep{\fill} } l c c } \hline \hline
    & $A\Omega_{\rm{eff}}$   &  Accumulative  \\
    & [km$^2$sr] & factor \\ \hline
Testbed analysis & 1.5E-4 & 1 \\ 
Testbed trigger& 1.5E-3 & 10 \\
ARA one-station trigger& 4.1E-3 & 28\\
ARA37 trigger & 1.3E-1& 900 \\ \hline \hline
\end{tabular*}
\caption{\label{tab:toara37} Factors that bring the Testbed sensitivity to ARA37 sensitivity for $E_{\nu}= 10^{18}$~eV using AraSim.}
\end{table}

In addition to the improvements in sensitivity from increasing the number of deep stations as listed in Table~\ref{tab:toara37}, we expect that our analysis efficiencies will improve in the next neutrino searches in the deep stations.
For example, the SPS Geometric Cut can be redefined for the deep stations due to the increased distance from the SPS and IceCube.
Whereas the Testbed views the SP region (defined by IceCube) as $34^{\circ}$ wide in azimuth, A1 views the SP region as $21^{\circ}$ and A2 and A3 each see it as 15$^{\circ}$.
Therefore, this same cut would eliminate 10\% of events in A2 and A3 instead of the 20\% in the Testbed unless the cut is eliminated completely, which is our aim.

Going from the Testbed to a design station, the directional reconstruction is expected to improve because the number of borehole antennas for each polarization increases from 4 to 8 and thus the number of pairs that go into the interferometric map increases from 6 to 28.
This will further reduce the rate of misreconstructions to the ice, below the estimated $<$1\% for the Testbed as described in Section~\ref{sec:EAS}, and reduce or eliminate the need for the geometric cuts on repeating locations.

Likewise, we aim to remove the IceCube Drilling Season Cut, which had removed 30\% of each year?s data set, since the 2011-2012 data-taking period was the last season where IceCube was actively drilling and the SPS will have been quieter in subsequent years when the deep stations were deployed.
The Delta Delay Cut is redundant with other cuts in this analysis (see Table~\ref{tab:events}) and can be removed.
Also, the Saturation Cut will have less of an effect on future analyses because of the larger volume of ice seen by the deep stations.
Due to the shadowing effect (see Fig.~\ref{fig:RayTrace}), the Testbed can only view events up to 3~km away, while the deep stations can view events out to $\sim$9~km.
Thus, a smaller fraction of events in the deep stations will be strong enough to saturate.
Based on a simulation of $10^{18}$ eV neutrinos, 25\% of triggered events in the Testbed will saturate compared to 12.5\% for a deep station.

The Calibration Pulser Timing Cut window can be significantly narrowed since newer stations have a more stable calibration pulser generation module than the Testbed.
The Calibration Pulser Timing Cut window for the Interferometric Map Analysis was conservatively set at 80~ms around the readout time resulting in an 8\% loss in efficiency.
Following an analysis of the rubidium clock timing, the Coherently Summed Waveform Analysis used a $\sim$100~ns window, and this much reduced window will be used in all future analyses.

We will also be able to remove the Gradient Cut if the background that this cut was designed to reject is not seen in the deep stations.
Lastly, the Good Baseline Cut can be removed with an improved filtering/rejection technique in progress.

Removing or loosening the cuts as described above results in an analysis efficiency of 35\% at a neutrino energy of $10^{18}$~eV, which is approximately a factor of four improvement compared to this analysis.
This does not account for additional improvements due to the increased number of antennas at greater depth and improved electronics.
We also expect a reduction in backgrounds due to increased distance from sources (e.g. SPS) for deep stations.

We also expect the trigger level sensitivity to be improved for the deep stations.
With the larger number of required coincidences in going from the Testbed to the deep stations (from 3/8 borehole antennas in the Testbed to 4/8 in each polarization of a deep station), we expect to reduce the trigger threshold by 15\% in power, keeping the same global trigger rate at 100~Hz and the coincidence window at 110~ns.

\section{Acknowledgements}
We thank the National Science Foundation for their generous support through Grant NSF OPP-1002483 and Grant NSF OPP-1359535, Taiwan National Science Councils Vanguard Program: NSC 102-2628-M-002-010 and the the FRSFNRS (Belgium). 
A. Connolly would also like to thank the National Science Foundation for their support through CAREER award 1255557, and also the 
Ohio Supercomputer Center. 
K. Hoffman would likewise like to thank the National Science Foundation for their support through CAREER award 0847658.
A. Connolly, H. Landsman and D. Besson would like to thank the United States-Israel Binational Science Foundation for their support through Grant 2012077.
A. Connolly, A. Karle and J. Kelley would like to thank the National Science Foundation for the support through BIGDATA Grant 1250720.
We are grateful to the U.S. National Science Foundation-Office of Polar Programs and the U.S. National Science Foundation-Physics Division.
We would also like to thank the University of Wisconsin Alumni Research Foundation, the University of Maryland and the Ohio State University for their support. 
We would also like to thank Raytheon Polar Services Corporation and the Antarctic Support Contractor, Lockheed, for field support.
We thank Chris Weaver from the University of Wisconsin for his work developing the RaySolver algorithm  used in AraSim, and David Saltzberg
for helpful conversations.

\appendix
\section { Appendix A: First estimation on Ray Tracing } \label{sec:appen_raytracing}

RaySolver in AraSim uses a semi-analytic approach to obtain the first estimated launching angle at the source location
given the horizontal distance to the source and indices of refraction at the source and target.
Here, we provide the derivation of Eq.~\ref{eq:raytracing} for the exponential index of refraction model~\cite{Sodha:1967}.
The index of refraction model is given as:
\begin{equation}
n(z) = A + Be^{C\cdot z} \label{eq:appen_1}
\end{equation}
where $n$ is index of refraction value, $z$ is the depth, and $A$, $B$, and $C$ are fitted parameter values from the South Pole measurements.
From the above equation, we can obtain:
\begin{align}
\frac{dn}{dz} &= B C  e^{C\cdot z} \nonumber \\
\frac{dn}{n} &= \frac{B C e^{C \cdot z}}{A+B e^{C\cdot z}} dz, \label{eq:appen_2}.
\end{align}

Now we take $\theta$ to be the launch angle of a signal with respect to the normal to the surface.   Taking $n_{\rm{r}}$ and $\sin{\theta}$ for the index of refraction and refracted angle with respect to normal, we can derive using Snell's law:
\begin{align}
n\sin{\theta} &= n_{\rm r} \sin{\theta_{\rm{r}}} \nonumber \\
dn &= \frac{-n_{\rm{r}} \sin{\theta_{\rm{r}}}}{\sin^2{\theta}} \cos{\theta} d\theta \nonumber \\
\frac{dn}{n} &= - \cot \theta d\theta \label{eq:appen_3}
\end{align}
where $\theta$ is the direction of the ray with respect to the direction normal to the surface.

Using Eq.~\ref{eq:appen_2} and \ref{eq:appen_3} and integrating from the source location to the target location, we obtain:
\begin{align}
\int_{z_0}^{z}{\frac{BCe^{Cz}}{A+Be^{Cz}}} & = \int_{\theta_0}^{\theta} {-\cot{\theta}}  \nonumber \\
\ln (A+Be^{C\cdot z'})\bigr|^{z}_{z_0} & = - \ln (\sin \theta ' )\bigr|^{\theta}_{\theta_0} \nonumber \\
\frac{A+Be^{C\cdot z}}{A+Be^{C\cdot z_0}} & = \frac{\sin \theta_0}{\sin \theta} \nonumber \\
\theta = \arcsin & \left( \sin \theta_0 \frac{A+Be^{C\cdot z_0}}{A+Be^{C\cdot z}} \right). \label{eq:appen_4}
\end{align}
where $\theta_0$ is the launch angle at the source location with respect to normal to the surface and $z_0$ is the depth at the source.

Using Eq.~\ref{eq:appen_4} and the relation $dx/dz~=~\tan \theta$:
\begin{align}
dx = \tan \left[ \arcsin \left( \sin \theta_0 \frac{A+Be^{C\cdot z_0}}{A+Be^{C\cdot z}} \right) \right] dz \label{eq:appen_5}
\end{align}
and integrating the equation from the source location to the target location, we obtain:
\begin{align}
x-x_0 & = \int ^z_{z_0} \tan \left[ \arcsin \left( \sin \theta_0 \frac{A+Be^{C\cdot z_0}}{A+Be^{C\cdot z'}} \right) \right] dz' \nonumber \\
 & = \int ^z_{z_0} \frac{ \sin \theta_0 ( A+Be^{C\cdot z_0})}{ A+Be^{C\cdot z'} \sqrt{ 1- \left( \frac{ \sin \theta_0 (A+Be^{C\cdot z_0})}{A+Be^{C\cdot z'}} \right)^2 } } dz'. \label{eq:appen_6}
\end{align}

In order to make above Eq.\ref{eq:appen_6} more manageable, we let:
\begin{align}
\sigma_0 & \equiv \sin \theta_0 \nonumber \\
n & \equiv  A+Be^{C\cdot z} \nonumber \\
dn & = BCe^{Cz} dz \nonumber \\
dz & = \frac{dn}{C(n-A)} \label{eq:appen_7}
\end{align}
and with these substitutions, Eq.~\ref{eq:appen_6} becomes:
\begin{align}
x-x_0 = \frac{\sigma_0 n_0}{C} \int ^{n}_{n_0} \frac{dn'}{n'(n'-A) \sqrt{ 1- \left( \frac{\sigma_0 n_0}{n'} \right)^2 } } \nonumber \\
\frac{C}{\sigma_0 n_0} (x-x_0) = \int ^{n}_{n_0} \frac{dn'}{(n'-A) \sqrt{ n'^2 - \sigma_0^2 n_0^2 }}. \label{eq:appen_8}
\end{align}

Now substituting $n$ with $m \equiv n - A$, the equation becomes:
\begin{align}
\frac{C}{\sigma_0 n_0} & (x-x_0) \nonumber \\
 = & \int ^{m}_{m_0} \frac{dm'}{m' \sqrt{ m'^2 +2Am'+A^2 - \sigma_0^2 n_0^2 }}. \label{eq:appen_9}
\end{align}

From Eq.~\ref{eq:appen_9}, we can make the equation more compact with an additional substitution $X \equiv a+bm'+cm'^2$ where $a= A^2 - \sigma_0^2 n_0^2$, $b = 2A$, and $c = 1$.
With the replacement, the equation is now:
\begin{align}
\frac{C}{\sigma_0 n_0} & (x-x_0) = \int ^{m}_{m_0} \frac{ dm' }{m' \sqrt{X}}. \label{eq:appen_10}
\end{align}

With the condition that $a \geq 0$ due to the fact that $A$ is the index of refraction value at the deep ice and $\sigma_0 = \sin \theta_0 \geq 0$, the integration of the right-hand-side of the Eq.~\ref{eq:appen_10} can be solved using~\cite{tableofintegrals}:
\begin{align}
\int \frac{dx}{x \sqrt{X}} = \frac{-1}{\sqrt{a}} \ln \left( \frac{2a+bx+2\sqrt{aX}}{x} \right) .\label{eq:appen_11}
\end{align}

After the integration from Eq.~\ref{eq:appen_9}, the equation becomes:
\begin{align}
& \,\,\, \frac{C\sqrt{A^2-\sigma_0^2 n_0^2}}{\sigma_0 n_0} (x_0 - x) \nonumber \\
& = \ln \left(  \frac{\sqrt{\left( n^2-\sigma_0^2n_0^2\right) \left( A^2-\sigma_0^2n_0^2 \right)} + An-\sigma_0^2 n_0^2}{n-A}\right) \nonumber \\
& - \ln \left(  \frac{\sqrt{\left( n_0^2-\sigma_0^2n_0^2\right) \left( A^2-\sigma_0^2n_0^2 \right)} + An_0-\sigma_0^2 n_0^2}{n_0-A}\right) .
\label{eq:appen_12}
\end{align}
This is Eq.\ref{eq:raytracing}, and all values are given parameters from the index of refraction model and the source and target locations except the launching angle $\sigma_0$ which is the single unknown.

 \bibliographystyle{elsarticle-num} 
  \bibliography{elsarticle-template-num}





\end{document}